\documentclass[pre,twocolumn,showpacs,superscriptaddress,aps]{revtex4}

\usepackage{amsfonts}
\usepackage{amsmath}
\usepackage{amssymb}
\usepackage{graphicx}
\usepackage{dcolumn}
\usepackage{times}
\usepackage{txfonts}
\usepackage{mathrsfs}
\usepackage{subfigure}

\begin{document}

\title{Ultrashort pulses and short-pulse equations in $(2+1)-$dimensions}
\author{Y. Shen}
\affiliation{Department of Mathematics and Statistics, University of Massachusetts,
Amherst MA 01003-4515, USA}
\author{N. Whitaker}
\affiliation{Department of Mathematics and Statistics, University of Massachusetts,
Amherst MA 01003-4515, USA}
\author{P.G.\ Kevrekidis}
\affiliation{Department of Mathematics and Statistics, University of Massachusetts,
Amherst MA 01003-4515, USA}
\author{N.L.\ Tsitsas}
\affiliation{Department of Informatics,
Aristotle University of Thessaloniki,
GR-54124 Thessaloniki, Greece}
\author{D.J.\ Frantzeskakis}
\affiliation{Department of Physics, University of Athens, Panepistimiopolis,
Zografos, Athens 15784, Greece}

\begin{abstract}
In this paper, we derive and study two versions of the short pulse equation (SPE) in $(2+1)-$dimensions. Using Maxwell's equations as a starting point, and suitable Kramers-Kronig formulas for the permittivity and permeability
of the medium, which are 
relevant, e.g., to left-handed metamaterials and dielectric slab waveguides, 
we employ a multiple scales technique to obtain the relevant models.
General properties of the resulting $(2+1)$-dimensional SPEs,
including fundamental conservation laws, as well as the Lagrangian and Hamiltonian structure
and numerical simulations for one- and two-dimensional
initial data, are presented. Ultrashort 1D breathers appear to be 
fairly robust, while rather general
two-dimensional localized initial conditions are transformed into
quasi-one-dimensional dispersing waveforms. 

\end{abstract}


\maketitle

\section{Introduction}

Ultrashort pulses, having a duration of a few optical cycles, have been the subject of intense study over
the last years; this is due to the fact that they find many applications in various contexts, ranging from
light-matter interactions, harmonic generation, attosecond physics, nonlinear optics, and others \cite{bk}.
A theme of particular interest related to ultrashort pulses, is their evolution in nonlinear media
characterized by an intensity-dependent refractive index. In that respect, models describing ultrashort
pulses in nonlinear media, as well as their systematic study, have attracted much attention;
see, e.g., Refs.~\cite{koz,mih1,besp,ls,mih2,mih3,mih4,LM}.
Prominent examples are the
modified Korteweg-de Vries (mKdV) equation \cite{mih1}, the sine-Gordon (sG) equation \cite{ls,mih2},
combined mKdV-sG equations \cite{mih3,mih4,LM} among
others. Note that most of these works refer
to the one-dimensional (1D) setting; in the two-dimensional (2D) one, pertinent studies are related
to few-cycle solitons described by the generalized Kadomtsev-Petviashvilli (KP) equation \cite{mih5} and
collapse dynamics of ultrashort spatiotemporal pulses \cite{LKM}.

Another model describing ultrashort pulse dynamics is the so-called {\it short-pulse
equation} (SPE), which was first derived in the context of nonlinear fiber optics \cite{sw} and later
in the context of nonlinear metamaterials \cite{tsitsas2}. From the physical point of view, the
interest in the SPE model arises from the fact that its few-cycle pulse solutions have been shown to compare more
favorably to the ones of the original Maxwell's equations, as compared to pertinent solutions of
the more traditional nonlinear Schr\"{o}dinger (NLS) model \cite{sw,sw2}. Furthermore, this model is also
interesting from a mathematical point of view, due to the existence of an infinite hierarchy
of conserved quantities \cite{brun}, its connection to the sG model and, thus, to its complete
integrability \cite{saksak1}. The SPE admits various types of solutions, including
singular soliton solutions -- the so called loop solitons \cite{saksak2} -- as well as other
non-singular solutions, such as peakons, breather- and periodic-type waveforms \cite{tsitsas2,saksak2,spesol,mats}.
Note that, recently, wave-breaking phenomena \cite{liupel}, as well as the global
well-posedness question \cite{pelsak} of the SPE were also investigated.
The above volume of work refers to the 1D setting; to the best of our knowledge, the SPE has not been
considered or analyzed so far in 2D.

In this work, we derive and study two different versions of the SPE in $(2+1)-$dimensions, namely the SPE-I and SPE-II.
In particular, starting from Maxwell's equations, and assuming general Kramers-Kronig or Sellmeier formulas for the permittivity and permeability (see, e.g., Ref.~\cite{skob}), we use a multi-scale expansion method
to obtain the SPE-I and SPE-II models. Then, we study the general properties of each model,
present the Hamiltonian, Lagrangian and momenta, and also obtain zero-mass constraints that are used
in the simulations (and, specifically, in the preparation of the initial data). Next, we explore the dynamics
of ultrashort pulses in 2D, employing as initial conditions, either the breather solution of the underlying 1D SPE or
a waveform localized in 2D; our purpose is to investigate if ultrashort pulses are prone to transverse instabilities (induced by the presence of diffraction in our models) and also identify purely 2D structures that can be supported by
the SPE-I and SPE-II. We find that the 1D breathers are stable in the 2D setting, while 2D initial data are gradually transformed into quasi-1D waveforms, reminiscent of the 1D solutions.

Our presentation is structured as follows. In Sec.~II, we derive the SPE-I and SPE-II models.
Sections III and IV are devoted to the general properties and numerical study of SPE-I and SPE-II, respectively. Finally, in Sec.~V, we summarize and discuss our conclusions.

\section{Derivation of 2D short-pulse equations}

We consider the propagation of $TE_z$ (transverse-electric field propagating along the $z$-axis) electromagnetic
(EM) waves in a planar metamaterial or optical waveguide structure.
In particular, we consider the case where
the electric and magnetic field components take the form ${\bf E}(x,z,t)={\bf \hat{y}} E_y(x,z,t)$ and ${\bf H}(x,z,t)={\bf \hat{x}}H_x(x,z,t)+{\bf \hat{z}}H_z(x,z,t)$,
where ${\bf \hat{x}}$, ${\bf \hat{y}}$, ${\bf \hat{z}}$ are the unit vectors along the $x$, $y$, $z$ directions,
respectively, and we have assumed no variations (i.e., a homogeneous
medium) with respect to the variable $y$. Under these assumptions, we may use
Maxwell's equations -- namely, Amp\'{e}re's and Faraday's laws --
take, respectively, the following form:
\begin{eqnarray}
&&-\frac{\partial H_z}{\partial x} + \frac{\partial H_x}{\partial z} =\frac{\partial D_y}{\partial t},
\label{A} \\
&&\frac{\partial E_y}{\partial z} =\frac{\partial B_x}{\partial t},
\qquad
\frac{\partial E_y}{\partial x} =-\frac{\partial B_z}{\partial t}.
\label{F}
\end{eqnarray}
Here $D_y$ is the $y$-component of the displacement vector ${\bf D}={\bf \hat{y}}D_y$.
Furthermore, we assume that the magnetic induction
vector ${\bf B}$ is connected with the magnetic field intensity ${\bf H}$
by means of the constitutive relation ${\bf \hat{B}} = \hat{\mu}(\omega) {\bf \hat{H}}$, where
$\hat{\mu}(\omega)$ is the linear magnetic permeability (hereafter, we use $f$ and $\hat{f}$ to denote
any function $f$ in the time- and frequency-domain, respectively). Additionally,
we assume that the considered structure exhibits a weak cubic (Kerr-type) nonlinearity in its dielectric response.
In other words, $D_y=\epsilon \ast E_y + P_{NL}$, where $\epsilon$ is the permittivity,
$\ast$ denotes the convolution integral $f(t)\ast g(t)=\int_{-\infty}^{+\infty}f(\tau)g(t-\tau)d\tau$ of any functions $f(t)$ and $g(t)$, while the nonlinear polarization $P_{NL}$ is of the form,
\begin{eqnarray}
P_{NL}&=&\epsilon_{0} \int_{-\infty}^{+\infty} \chi_{NL}(t-\tau_1, t-\tau_2, t-\tau_3)
\nonumber \\
&\times& E_y(\tau_1) E_y(\tau_2) E_y(\tau_3) d\tau_1 d\tau_2 d\tau_3.
\label{pnl}
\end{eqnarray}
Here, $\epsilon_{0}$ is the dielectric constant of vacuum and $\chi_{NL}$ is the nonlinear electric
susceptibility of the medium. In the case of small-amplitude, {\it ultra-short} pulse propagation,
the nonlinear response can safely be considered to be {\it instantaneous}, namely,
\begin{equation}
\chi_{NL}(t-\tau_1, t-\tau_2, t-\tau_3)
= \kappa \delta(t-\tau_1)\delta(t-\tau_2)\delta(t-\tau_3),
\label{xnl}
\end{equation}
where $\kappa$ is the Kerr coefficient given by $\kappa = \pm E_{c}^{-2}$, with $E_c$ being
a characteristic electric field value;
generally, both cases of focusing ($\kappa>0$) and defocusing ($\kappa<0$) dielectrics are possible.
Notice that Eqs.~(\ref{pnl}) and (\ref{xnl}) imply that
$P_{NL}=\epsilon_0 \kappa E_y^3$ and, thus, $D_y = \epsilon \ast E_y + \epsilon_0 \kappa E_y^3$. Substituting
the considered form of the constitutive relations into Eqs.~(\ref{A})-(\ref{F}), we derive the following
equation for the $y$-component of the electric field intensity ($E_y$) which, for convenience,
will be denoted hereafter by $E$:
\begin{equation}
\nabla^2 E -\partial_t^2 (\epsilon \ast \mu \ast E) - \epsilon_0 \kappa \partial_t^2 (\mu \ast E^3)=0,
\label{Max}
\end{equation}
where $\nabla^2 \equiv \partial_x^2+ \partial_z^2$ is the Laplacian in the $(x, z)$-plane.

Equation~(\ref{Max}) is the $(2+1)$-dimensional generalization of the 1D Klein-Gordon type
model derived in the context of nonlinear fiber optics \cite{sw,jones} (in this case, $\mu = {\rm const.}$) and nonlinear metamaterials \cite{longhi,tsitsas2} (in this case, $\partial_\omega \hat{\mu} \ne 0$). Below, we will analyze the
latter (more general) case, and assume that both permittivity and permeability are frequency dependent. In particular,
considering general Kramers-Kronig or Sellmeier formulas (see, e.g., Ref.~\cite{skob}), we assume that the frequency dependence of $\hat{\epsilon} \equiv \hat{\epsilon}(\omega)$ and $\hat{\mu} \equiv \hat{\mu}(\omega)$ can be approximated by the relations
\begin{equation}
\hat{\epsilon}(\omega) \approx \epsilon_0 \left(\alpha_1 - \frac{\alpha_2}{\omega^2} \right), \qquad
\hat{\mu}(\omega) \approx \mu_0 \left(\beta_1 - \frac{\beta_2}{\omega^2}\right),
\label{KK}
\end{equation}
where $\alpha_1$, $\alpha_2$, $\beta_1$ and $\beta_2$ are some constants.
The above approximations can be applied 
to the contexts of nonlinear left-handed metamaterials and 
nonlinear optical slab waveguides. Specifically, in the context of nonlinear left-handed metamaterials,
$\alpha_1 =1$, $\alpha_2 = \omega_p^2$, $\beta_1 = 1-F$ and $\beta_2 = F \omega_{\rm res}^2$,
where $\omega_p$, $F$ and $\omega_{\rm res}$ denote, respectively, the plasma frequency,
the filling factor, and the magnetic permeability resonance frequency \cite{tsitsas2}.
On the other hand, in the context of nonlinear optical slab waveguides, 
$\alpha_1 =\epsilon_r^{(0)}$, $\alpha_2 = \epsilon_r^{(2)}$, $\beta_1 = 1$ and $\beta_2 = 0$,
where $\epsilon_r^{(0)}$ and $\epsilon_r^{(2)}$ are relative dielectric constants
(with $\epsilon_r^{(2)}$ being measured in units of squared angular frequency) 
obtained by matching 
the full form of the permittivity with the first of Eqs.~(\ref{KK}) over a specific wavelength range
in the infrared regime \cite{sw}.

Next, we express Eq.~(\ref{Max}) in the frequency domain and substitute Eqs.~(\ref{KK}),
keeping terms up to order $\mathcal{O}(\omega^{-2})$ (i.e., assuming that $\alpha_1 \beta_2/\omega^4 \ll1$); then, expressing the 
resulting equation back in time domain, and measuring time, space, and field intensity $E^2$ in units of
$1/\sqrt{\alpha_2}$, $c/\sqrt{\alpha_1 \alpha_2 \beta_1}$ and $|\kappa|^{-1}$ respectively, we reduce
Eq.~(\ref{Max}) in the following dimensionless form:
\begin{equation}
\nabla^2 E -\partial_t^2 E - \alpha E - s_{\kappa} (\beta E^3 + \gamma \partial_t^2 E^3)=0.
\label{KG}
\end{equation}
In the above equation, $s_{\kappa}={\rm sign}(\kappa)$, while the other constants are given by:
\begin{equation}
\alpha = \frac{1}{\alpha_1} + \frac{\beta_2}{\alpha_2 \beta_1},
\quad
\beta = \frac{\beta_2}{\alpha_1 \alpha_2 \beta_1},
\quad
\gamma = \frac{1}{\alpha_1},
\label{constants}
\end{equation}
Note that in the context of nonlinear metamaterials $\alpha=1+\beta=1+F\omega_{\rm res}^2/[(1-F)\omega_p^2]$ and $\gamma=1$,
while in the context of nonlinear optical slab waveguides $\alpha=\gamma=1/\epsilon_r^{(0)}$ and $\beta=0$.
%

We now consider propagation of ultrashort pulses, of width $\varepsilon$, where $0<\varepsilon \ll 1$ is a formal small parameter, which will also set the field amplitude (see below). Then, we employ the method of multiple scales to derive from Eq.~(\ref{KG}) two different versions of short pulse equations in $(2+1)$-dimensions. In that regard, we introduce the following asymptotic expansion for the unknown field $E$:
\begin{equation}
E= \varepsilon E_1 (T, X_n, Z_n) + \varepsilon^{2} E_2 (T, X_n, Z_ n)+\ldots,
\label{an}
\end{equation}
where the functions $E_n$ depend on the spatial variables $X_n$ and $Z_n$ ($n=1,2,\ldots$),
as well as on the fast time variable $T$. Defining $Z_n$ as:
\begin{equation}
Z_n= \varepsilon^n z,
\label{zn}
\end{equation}
we consider two different definitions for $X_n$ and $T$, namely:
\begin{eqnarray}
X_n&=& \varepsilon^{n-1} x, \qquad T= \frac{t-z}{\varepsilon},
\label{I} \\
X_n&=& \varepsilon^{n} x, \qquad T= \frac{t-(x+z)/\sigma \sqrt{2}}{\varepsilon},
\label{II}
\end{eqnarray}
where $\sigma =\pm 1$. Then, substituting Eqs.~(\ref{an}), (\ref{zn}) and (\ref{I}) in Eq.~(\ref{KG}),
we derive at order $\mathcal{O}(\varepsilon)$
the following $(2+1)$-dimensional SPE for the unknown field $E_1$:
\begin{equation}
2 \frac{\partial^2 E_1}{\partial Z_1 \partial T} - \frac{\partial^2 E_1}{\partial X_1^2}
+\alpha E_1 +s_{\kappa} \gamma \frac{\partial^2 }{\partial T^2}(E_1^3) = 0.
\label{spe1}
\end{equation}
Equation~(\ref{spe1}) will be called hereafter SPE-I. Similarly,
substituting Eqs.~(\ref{an}), (\ref{zn}) and (\ref{II}) in Eq.~(\ref{KG}), we derive
[again at order $\mathcal{O}(\varepsilon)$] another $(2+1)$-dimensional version of the SPE, namely:
\begin{equation}
2\sigma \left(\frac{\partial^2 E_1}{\partial Z_1 \partial T}+\frac{\partial^2 E_1}{\partial X_1 \partial T} \right)
+\alpha E_1 +s_{\kappa} \gamma \frac{\partial^2 }{\partial T^2}(E_1^3) = 0,
\label{spe2}
\end{equation}
which will be called hereafter SPE-II. We note that variants of these models have been considered
in the past in the context of ultrashort propagation in nonlinear dielectrics \cite{koz,besp},
and more recently in relevant studies \cite{LM}, as well as in the context of collapse
in two-level media \cite{LKM}.


\section{The SPE-I: general properties and numerical study}

In this section, we focus on the SPE-I which, for simplicity of notation, is expressed in the form:
\begin{equation}
2E_{zt}-E_{xx}+\alpha E+s(E^3)_{tt}=0,
\label{2dspe_2}
\end{equation}
where $s= s_{\kappa} \gamma$ and subscripts denote partial derivatives, with $z$ being the evolution variable. Below, we will consider general properties of this equation and discuss its solutions.

\subsection{Properties and canonical structure}

First, we study the Hamiltonian structure of Eq.~(\ref{2dspe_2}). For this purpose, we
integrate Eq.~(\ref{2dspe_2}) with respect to time $t$ and, introducing the auxiliary field $E=\phi_t$,
we express Eq.~(\ref{2dspe_2}) as follows:
\begin{equation}
2\phi_{tz}-\phi_{xx}+\alpha\phi+s(\phi_t^3)_t=0.
\label{ve}
\end{equation}
Then, it can be verified that Eq.~(\ref{ve}) can be obtained from the variational principle, with Lagrangian density:
\begin{equation}
\mathcal{L}=-\frac{\alpha}{2}\phi^2+\frac{s}{4}\phi_t^4+\phi_t\phi_z-\frac{1}{2}\phi_x^2.
\end{equation}
From this Lagrangian density, we can derive the Hamiltonian:
\begin{eqnarray}
H&=&\int_{-\infty}^{+\infty} \int_{-\infty}^{+\infty}
\left( \frac{\partial\mathcal{L}}{\partial\phi_z}\phi_z-\mathcal{L} \right) {\rm d}t {\rm d}x
\nonumber \\
&=&\int_{-\infty}^{+\infty} \int_{-\infty}^{+\infty} \left( \frac{\alpha}{2}\phi^2-\frac{s}{4}\phi_t^4+\frac{1}{2}\phi_x^2 \right) {\rm d}t {\rm d}x,
\label{H}
\end{eqnarray}
as well as the momenta:
\begin{eqnarray}
\!\!\!\!\!\!\!\!\!\!
M_t&=&\int_{-\infty}^{+\infty} \int_{-\infty}^{+\infty}
\frac{\partial\mathcal{L}}{\partial\phi_z}\phi_t {\rm d}t {\rm d}x
=\int_{-\infty}^{+\infty}\int_{-\infty}^{+\infty} \phi_t^2{\rm d}t {\rm d}x,
\label{Mt} \\
\!\!\!\!\!\!\!\!\!\!
M_x&=&\int_{-\infty}^{+\infty}\int_{-\infty}^{+\infty} \frac{\partial\mathcal{L}}{\partial\phi_z}\phi_x{\rm d}t {\rm d}x
=\int_{-\infty}^{+\infty}\int_{-\infty}^{+\infty}\phi_t\phi_x{\rm d}t {\rm d}x. 
\label{Mx}
\end{eqnarray}

Let us next consider the Fourier transform of Eq.~(\ref{2dspe_2}) with respect to time $t$, which leads
to the equation:
\begin{equation}
(i \omega)\hat{E}_{z} =
\frac{1}{2}\hat{E}_{xx}-\frac{\alpha}{2}\hat{E}-\frac{s}{2}(i\omega)^2\hat{E^3}.
\label{2d-hat-spe}
\end{equation}
The above equation implies that
\begin{equation}
\hat{E}_{z}(\omega_t,x,z) =-\frac{i}{2}\bigg(
\frac{\hat{E}_{xx}-\alpha\hat{E}}{\omega}+s\omega \hat{E^3}\bigg)~,
\quad {\rm for}~~\omega \neq\ 0,
\end{equation}
\begin{equation}
\hat{E}_{xx}-\alpha\hat{E}=0, \quad {\rm for}~~\omega=0
\label{omegat=0}.
\end{equation}

In our numerical simulations below, we will set $\hat{E}(\omega=0,x,z)=0$, so that
Eq.~(\ref{omegat=0}) is satisfied. Thus, in this case, the Fourier transform of
the field $E$ leads to the following ``zero-mass constraint'':
\begin{equation}
\int_{-\infty}^{\infty} E(t,x,z) {\rm d}t=0 \quad {\rm for~~any}~~x,~z,
\label{constrain}
\end{equation}
which also holds for the traditional SPE in $(1+1)$-dimensions (see, e.g.,
the relevant analysis of Ref.~\cite{horikisjpa}).

\subsection{1D breather-like structures}

Let us now seek one-dimensional (1D) solutions of Eq.~(\ref{2dspe_2}),
by assuming that the unknown field $E$
depends on the traveling-wave coordinates $\xi$ and $\eta$, defined as:
\begin{equation}
\xi=z, \qquad \eta=t+cx+\frac{c^2}{2} z,
\label{1d_transform15}
\end{equation}
where $c$ is an arbitrary real constant setting the velocities of the traveling wave in the $(x,\,t)$
and $(x,\,z)$ planes.
Using the above variables, Eq.~(\ref{2dspe_2}) is reduced to the form:
\begin{equation}
2E_{\xi\eta}+\alpha E+s(E^3)_{\eta\eta}=0,
\label{tradspe}
\end{equation}
which is actually the $(1+1)$-dimensional SPE \cite{sw}. As shown in the simulations
of Ref.~\cite{spesol}, the most robust among the various solutions of the 1D SPE
is the breather-like structure (this solution satisfies the zero-mass constraint).
Naturally, this purely 1D structure satisfies
the full 2D SPE-I, Eq.~(\ref{2dspe_2}) and, thus, an interesting question concerns the
stability of this solution in the 2D space. A similar question appears in many physically relevant
models and, in many cases, the answer is that such ``planar'' solutions are prone to transverse
instabilities in higher-dimensional settings; as characteristic
examples, we mention the line soliton solutions
of the Kadomtsev-Petviashvilli-I (KP-I) equation which decay into lumps \cite{kuz1,inf}, or the
dark soliton stripes of the defocusing NLS equation which decay into
vortices \cite{kuz2,dep} in $(2+1)$-dimensions.

\begin{figure}[tbp]
\begin{center}
{\includegraphics[width=0.22\textwidth]{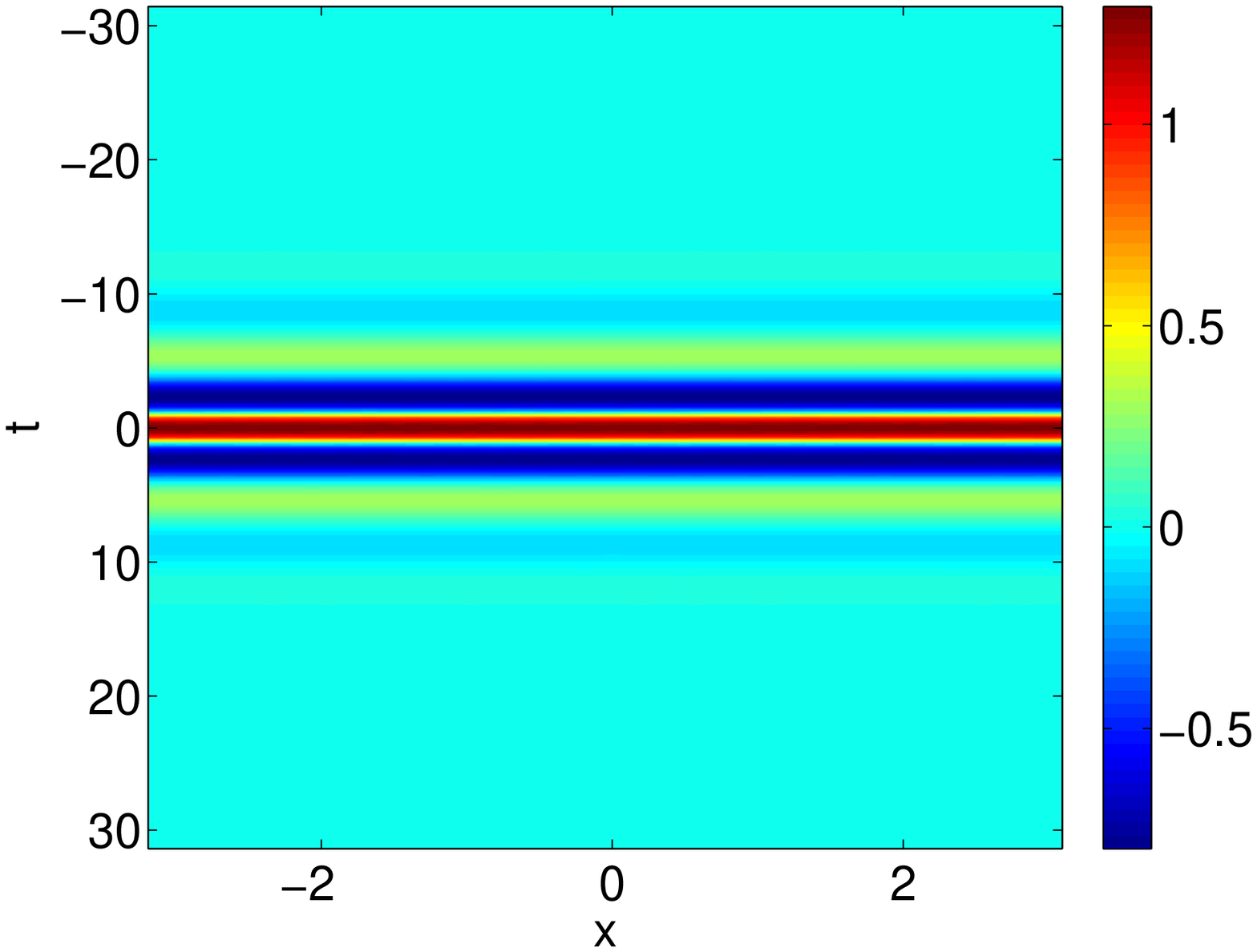}}
{\includegraphics[width=0.22\textwidth]{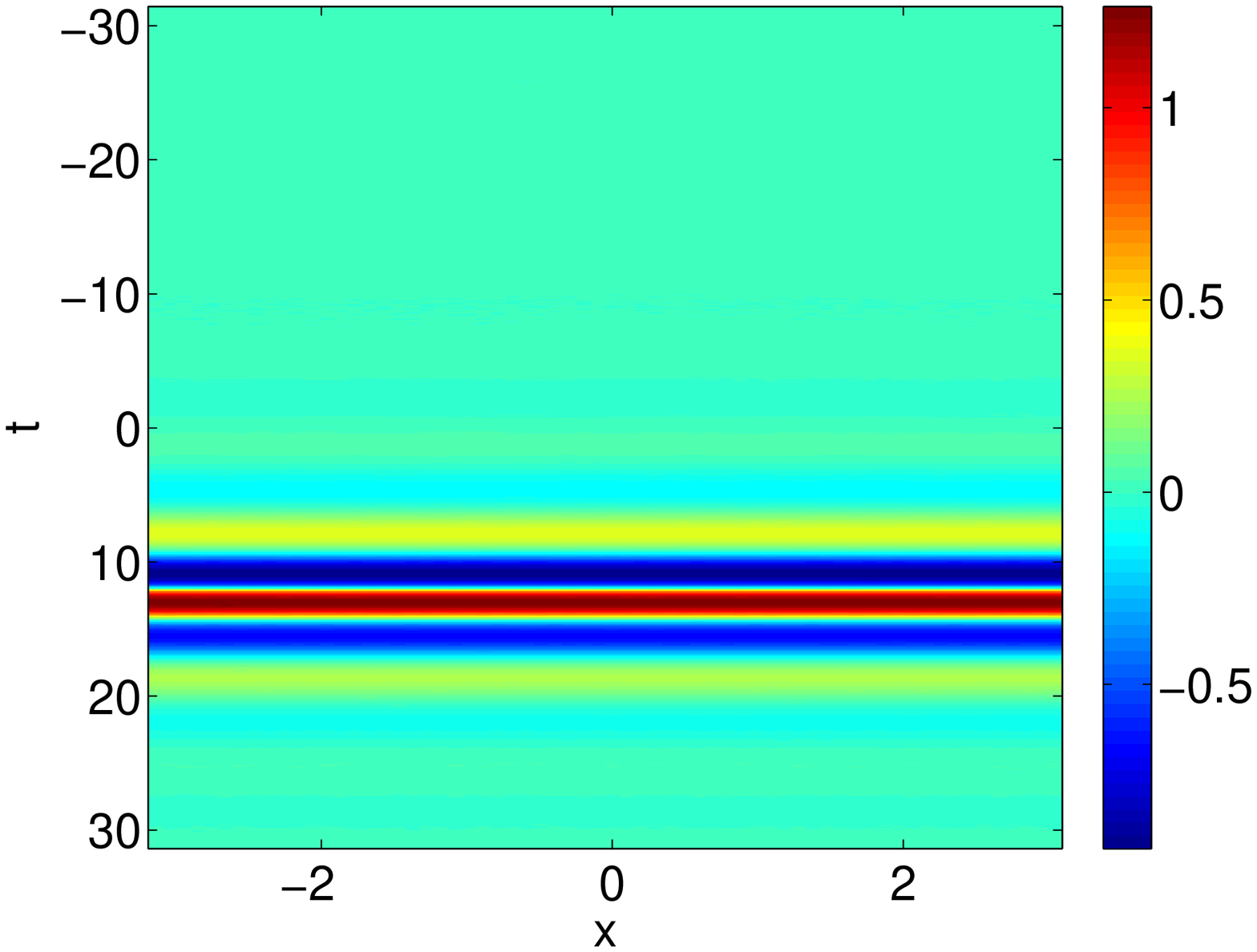}}
{\includegraphics[width=0.22\textwidth]{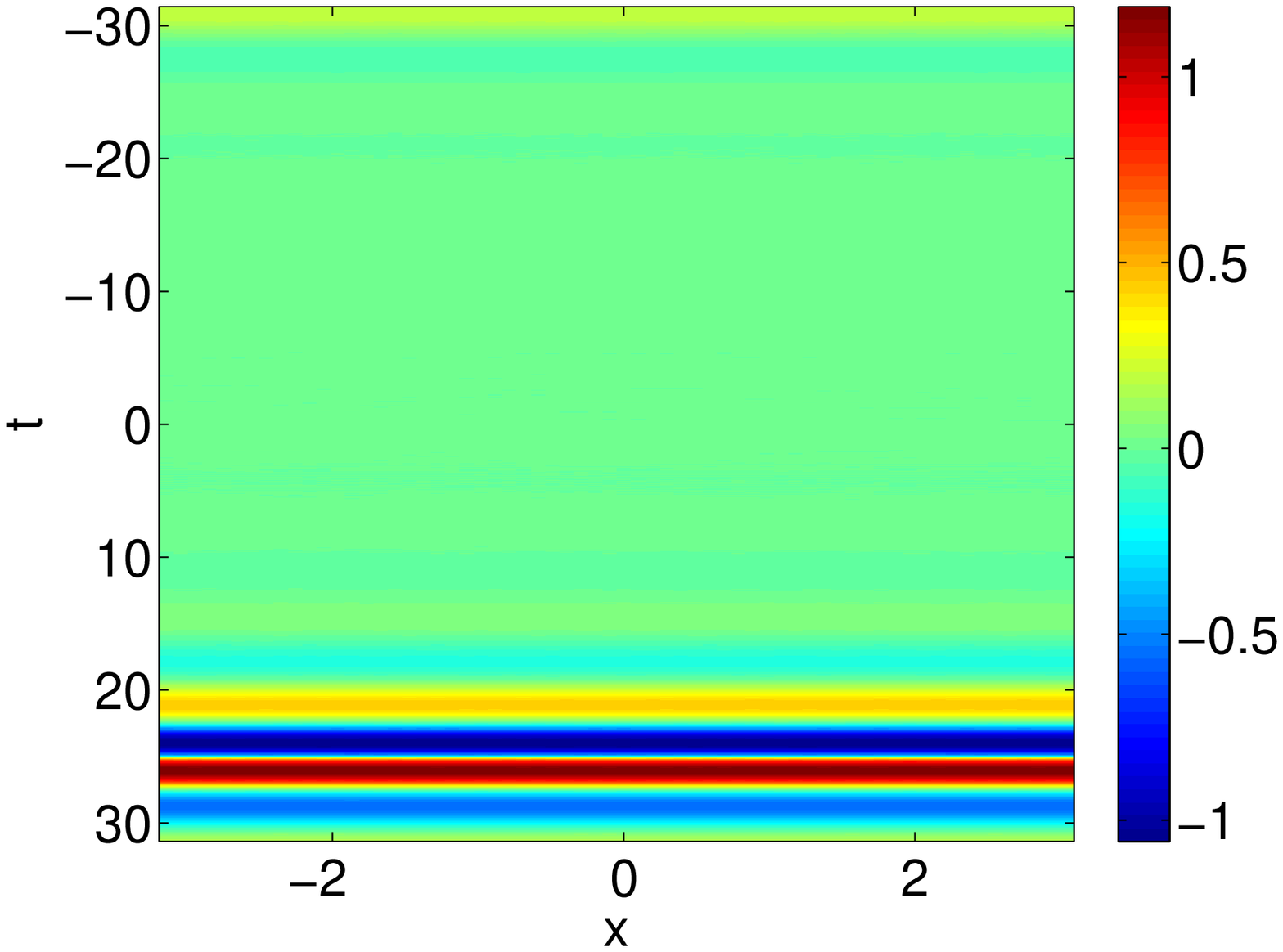}}
{\includegraphics[width=0.22\textwidth]{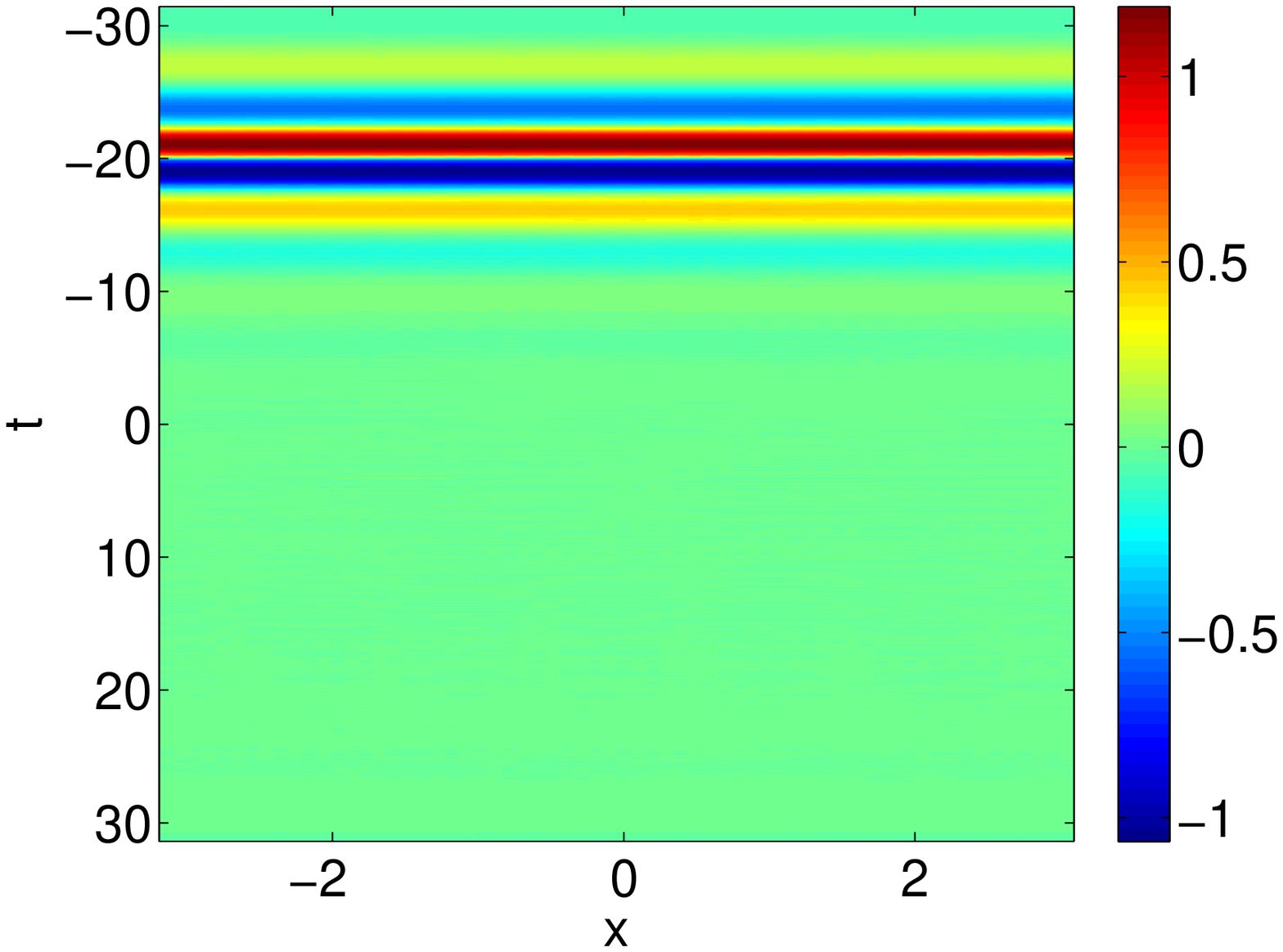}}
{\includegraphics[width=0.35\textwidth]{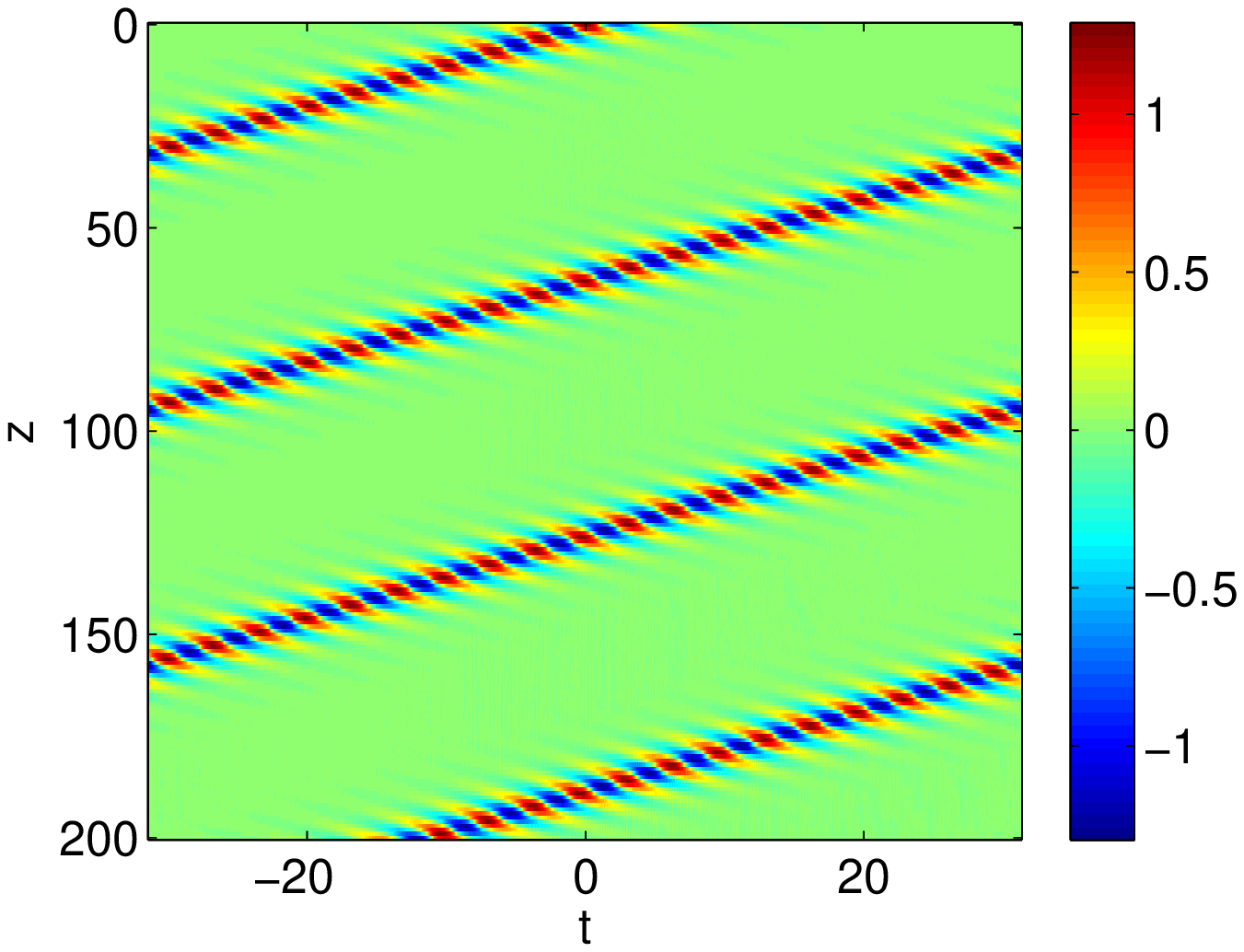}}
\end{center}
\caption{
(Color online)
Top (four) panels:  Contour plots showing profiles of the evolution
of a perturbed 1D breather in the $(x, t$) plane,
when evolved according to
Eq.~(\ref{2dspe_2}). Snapshots correspond to $z=0$ (top left), $z=50$ (top right)
$z=100$ (bottom left), and $z=150$ (bottom right).
Bottom panel: The evolution of the breather for $x=0$.
Parameter values are $c=0$, $s=-1/3$, and $\alpha = -2$.
}
\label{1Dperturb1}
\end{figure}
\begin{figure}[tbp]
  \begin{center}
  {\includegraphics[width=0.25\textwidth]{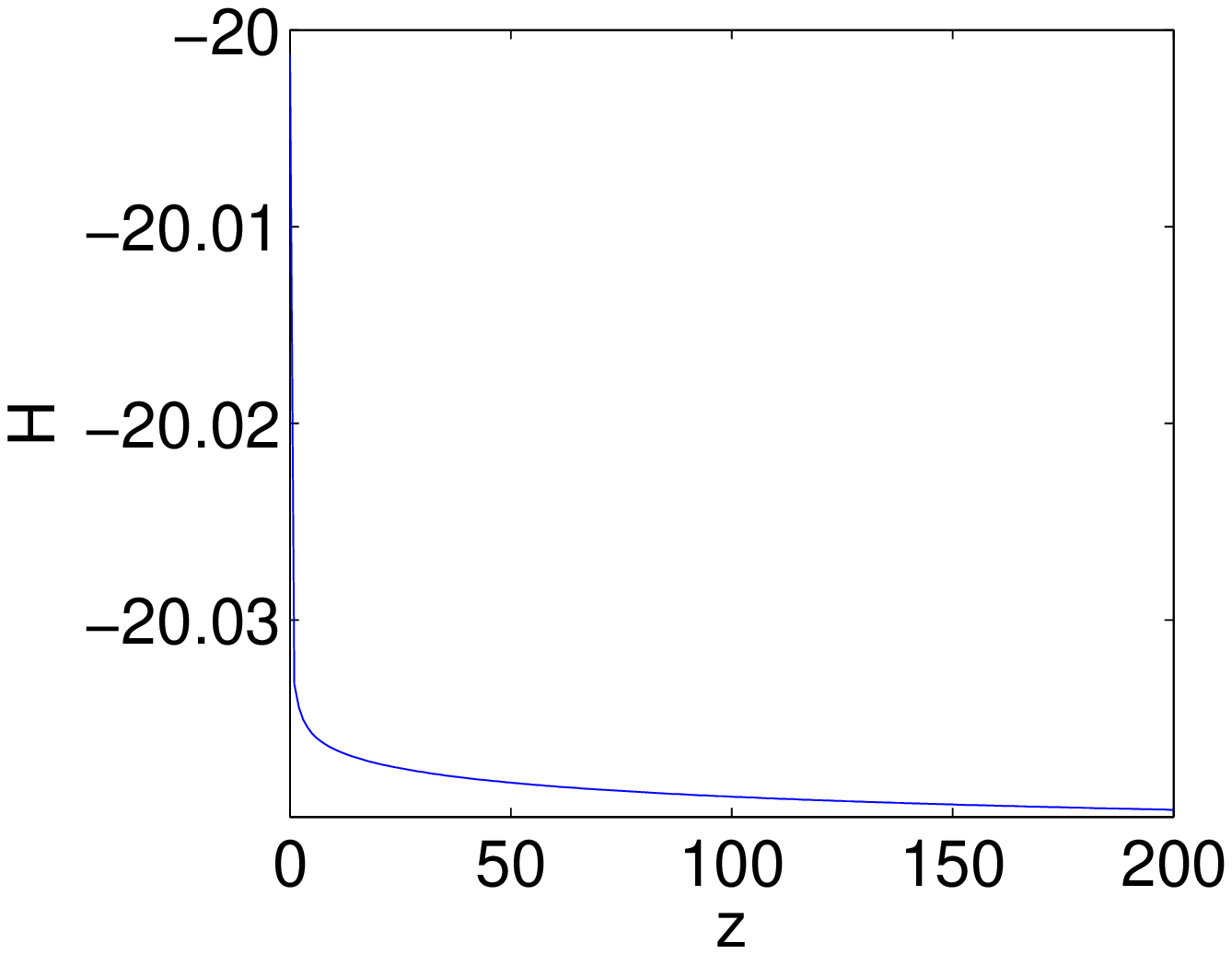}} \\
  {\includegraphics[width=0.23\textwidth]{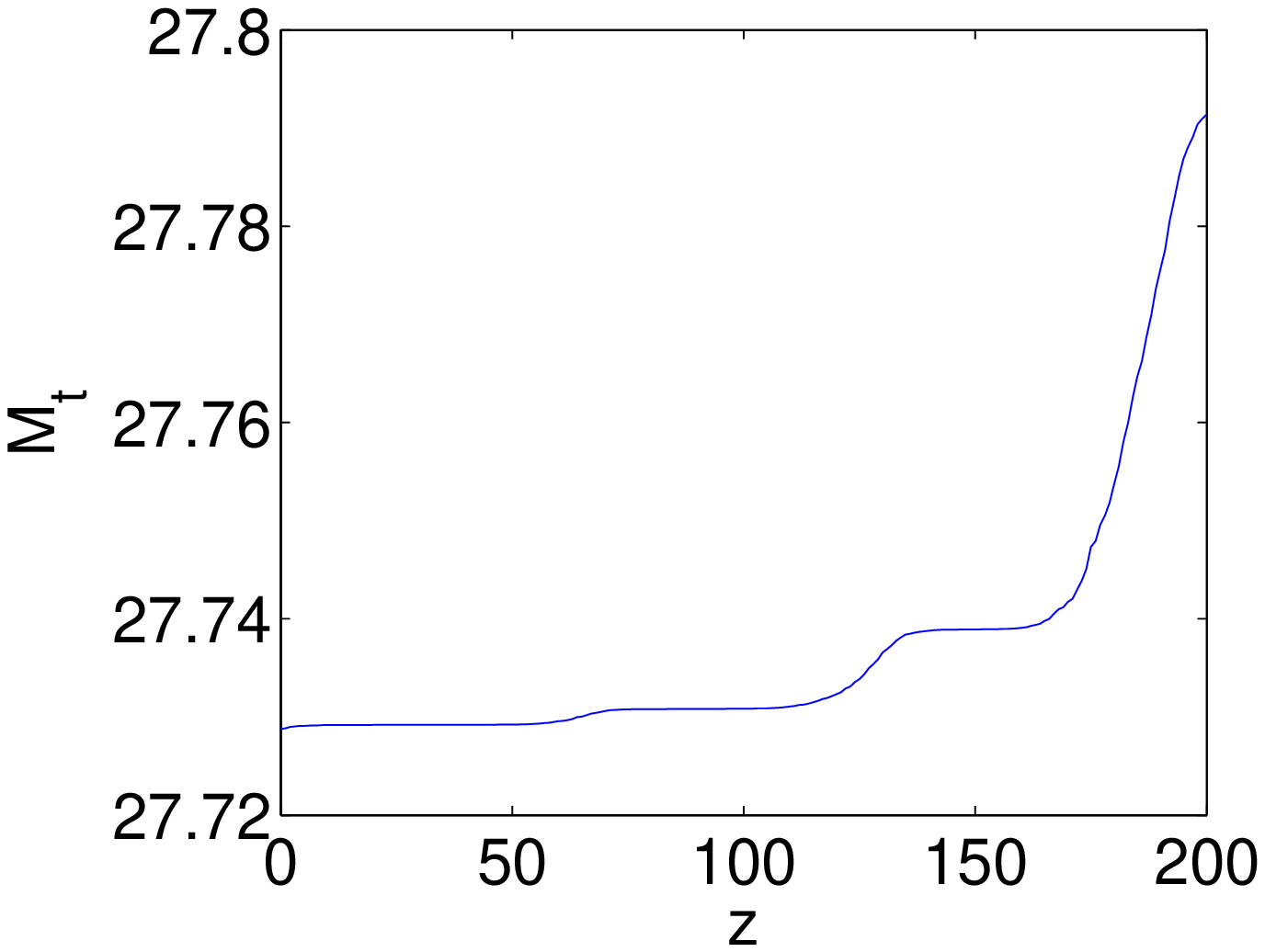}}
  {\includegraphics[width=0.22\textwidth]{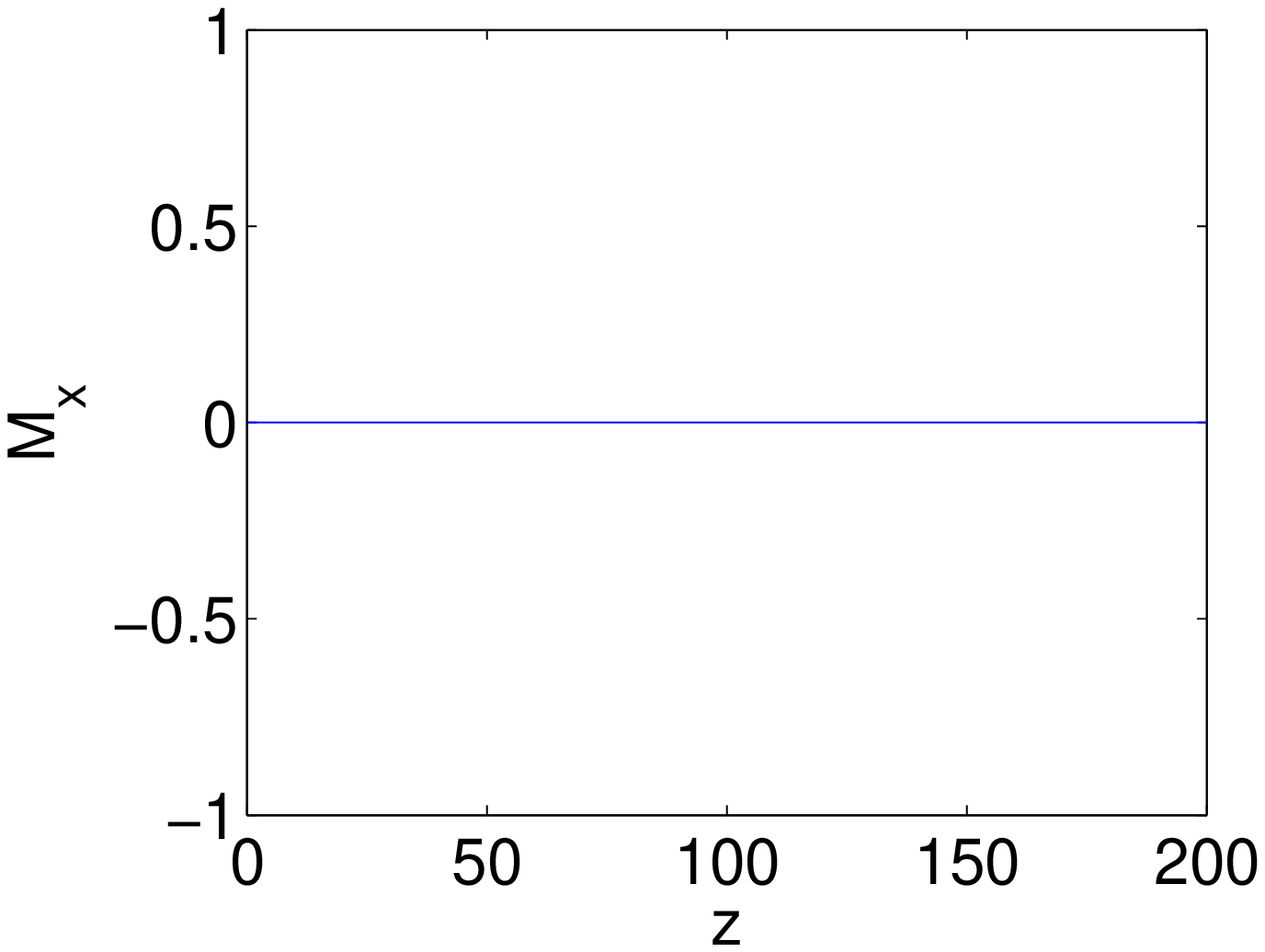}}
  \end{center}
\caption{(Color online) The evolution of the conserved quantities of the SPE-I for the simulation shown in Fig.~\ref{1Dperturb1}: Top panel shows the Hamiltonian $H$ and bottom panels show the momenta $M_t$ (left) and $M_t$ (right). The relative error for $H$ and $M_t$ is of order $10^{-4}$ and $10^{-3}$ respectively, and $M_x$ is zero.}
  \label{1Dperturb_hamitonian1}
\end{figure}

To study the stability of the 1D breather-like solution of Eq.~(\ref{tradspe}) in the framework of the
the full 2D SPE-I, we have used the following procedure. We employed the
breather solution of the 1D SPE Eq.~(\ref{tradspe}) and also added, as a perturbation, a small
noisy signal,
of amplitude of $1\%$ of the breather's amplitude. Then, using the resulting structure
(cf. top panel of Fig.~\ref{1Dperturb1}) as an initial condition, we numerically integrated
Eq.~(\ref{2dspe_2}) by means of a Galerkin method (and assuming periodic boundary conditions in
our numerical scheme). The results (corresponding to parameter values $c=0$, $s=-1/3$, and $\alpha = -2$)
are shown in the panels of Fig.~\ref{1Dperturb1}, in terms
of different contour plots depicting the profile of the 1D breather in the
$(x, t$) plane for various values of the propagation distance $z$
(and also the evolution at $x=0$ as a function of $(z,t)$). It is
clear that the breather is robust, at least up to $z=200$ (where the simulation ended).
We should also mention that for these simulations, we have also calculated the evolution
of the Hamiltonian and momenta [cf. Eqs.~(\ref{H})-(\ref{Mx})]. The results, depicted in
Fig.~\ref{1Dperturb_hamitonian1}, justify the conservation of these quantities with a
satisfactory (relative) accuracy, of order $10^{-3}$ or less.

We finally note that similar results (not shown here)
were also obtained for breathers with i.e.,
for $c\ne 0$ in Eq.~(\ref{1d_transform15}); in such a case,
the only difference is that the breather is ``tilted'', i.e.,
oblique in the ($x, t)$ plane with respect to
its direction in the case $c=0$ and follows a similar evolution (i.e., it is stable up to end of the simulation time).
%


\subsection{Localized initial data}

Having discussed the properties of the 1D breather in the 2D setting, we now turn our attention
to initial data associated with Eq.~(\ref{tradspe}), which are localized in both transverse directions,
$x$ and $t$. In that regard, it is convenient to consider at first the decomposition
$E(x,t)=f(t)g(x)$,
and substitute this ansatz in the zero-mass constraint, Eq.~(\ref{constrain}). This way, for nontrivial solutions, we
derive the necessary condition $\int_{-\infty}^{\infty}f(t)dt=0$.
%
%

\begin{figure}[tbp]
\begin{center}
{\includegraphics[width=0.22\textwidth]{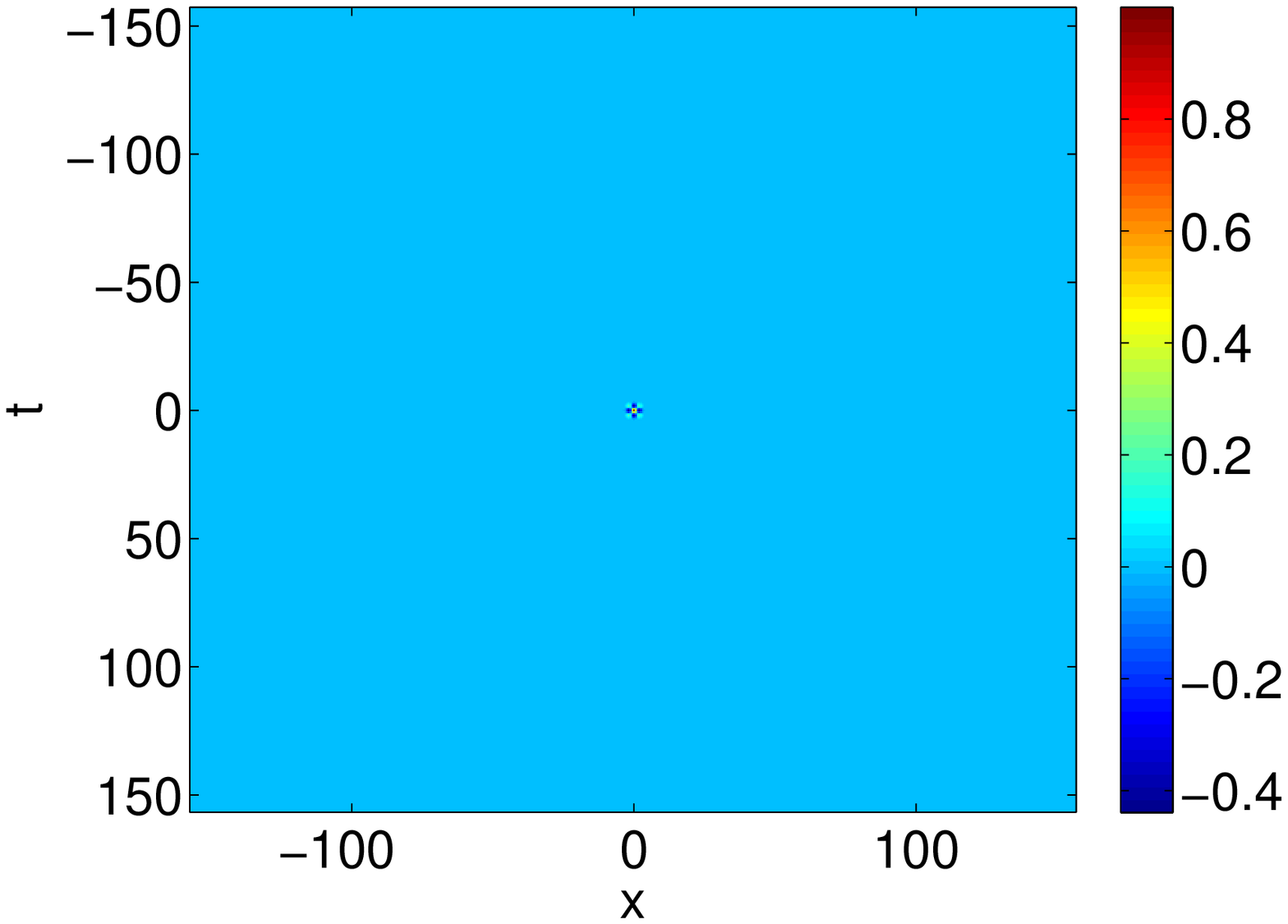}}
{\includegraphics[width=0.22\textwidth]{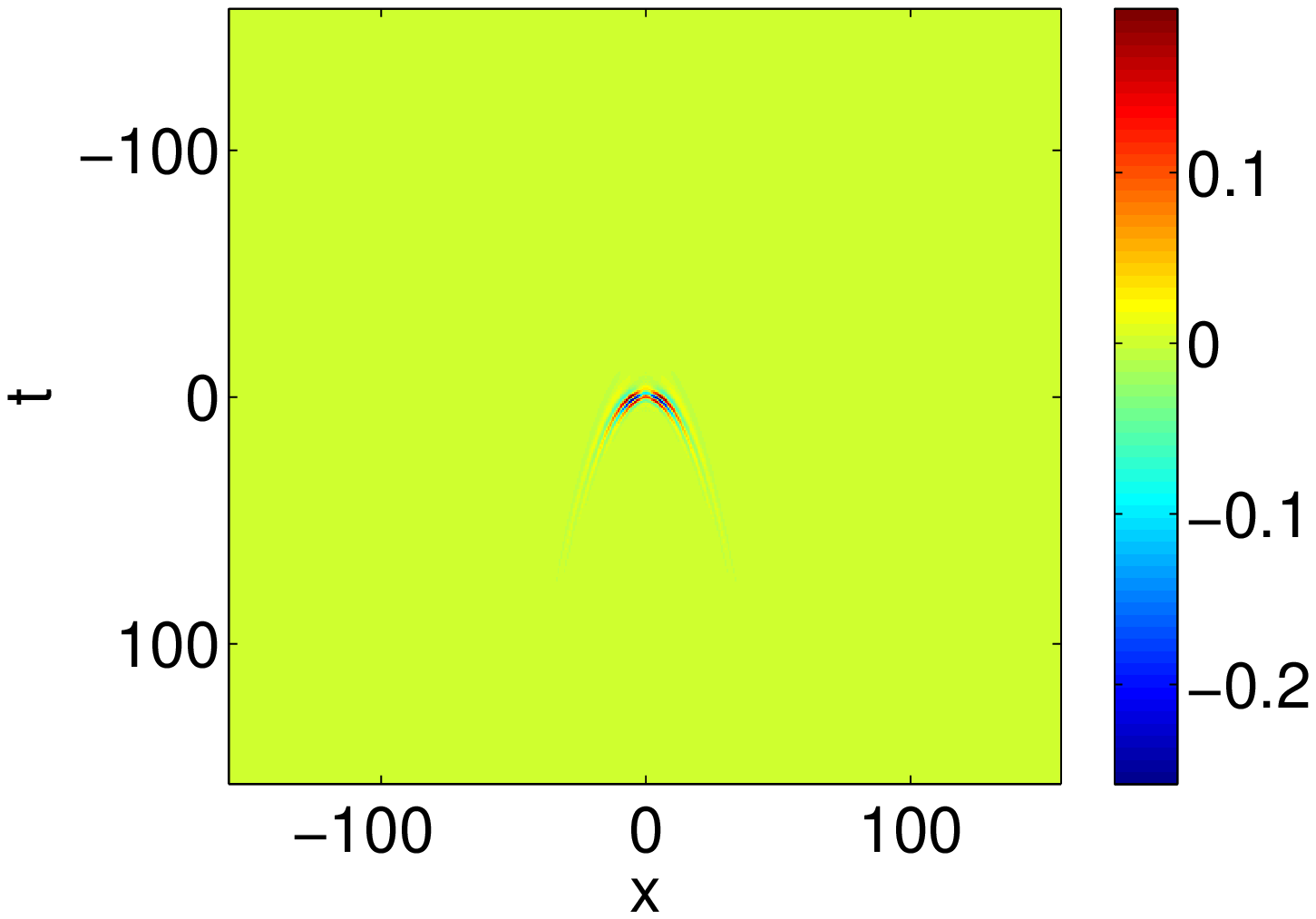}}
{\includegraphics[width=0.22\textwidth]{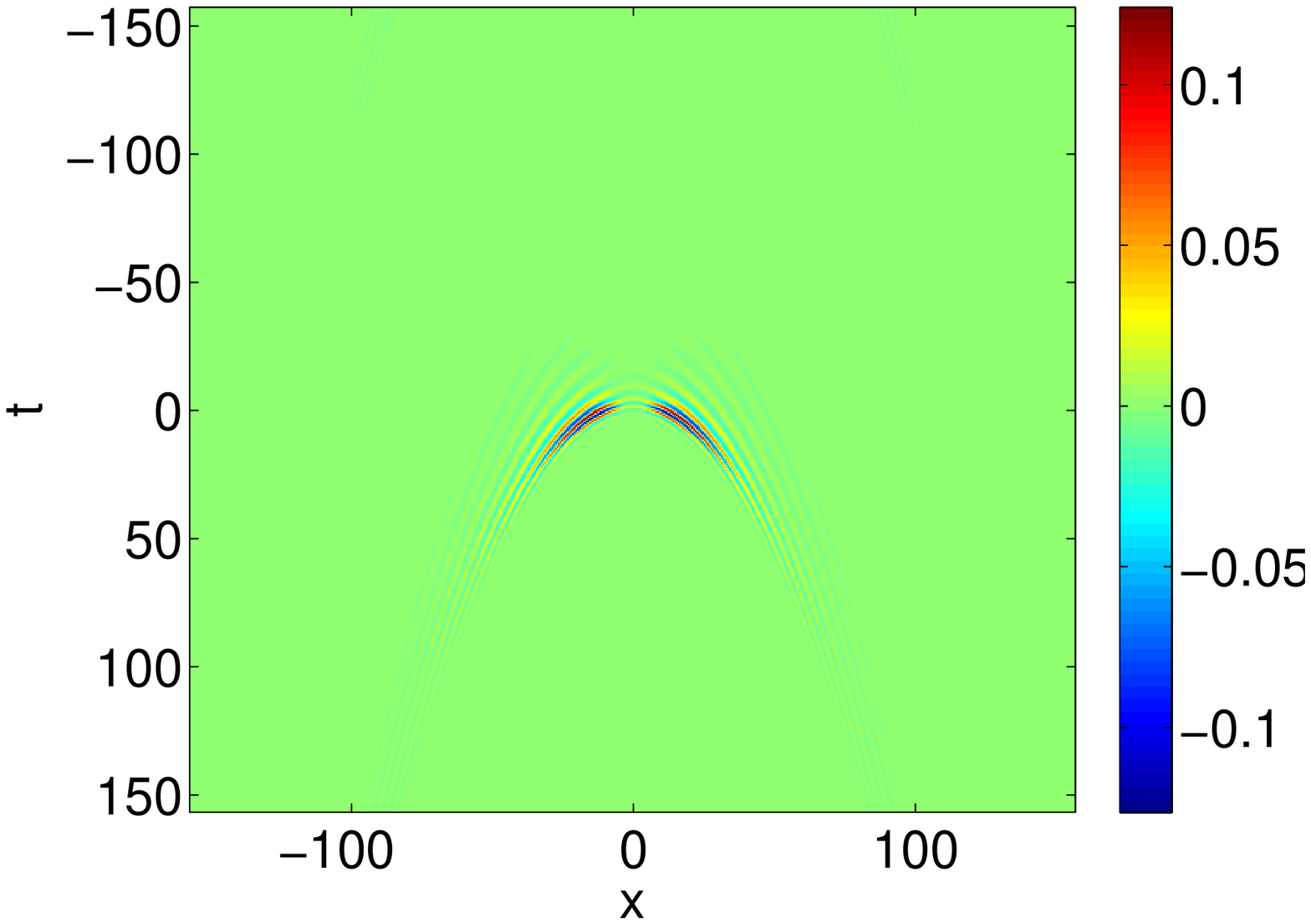}}
{\includegraphics[width=0.22\textwidth]{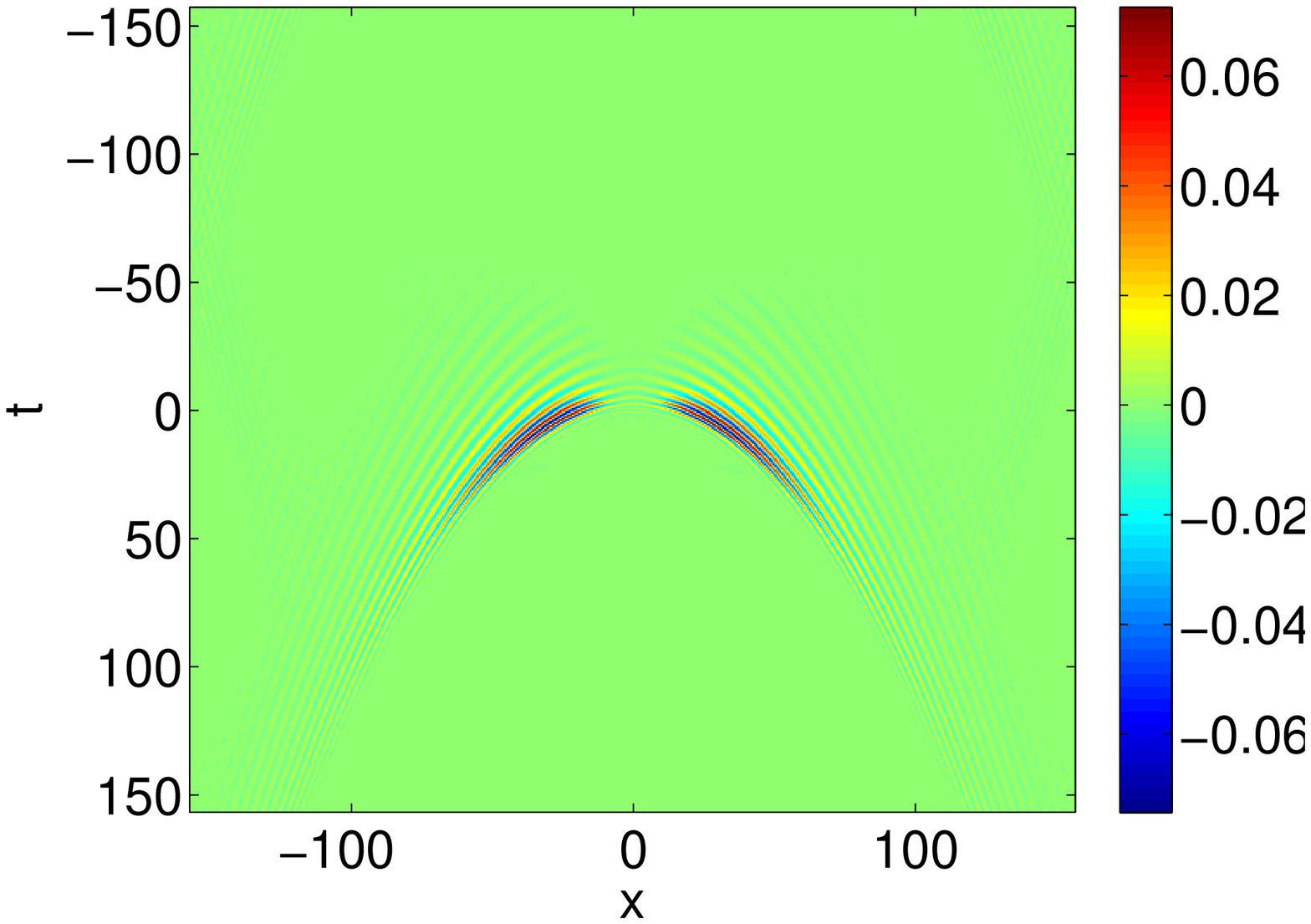}}
{\includegraphics[width=0.22\textwidth]{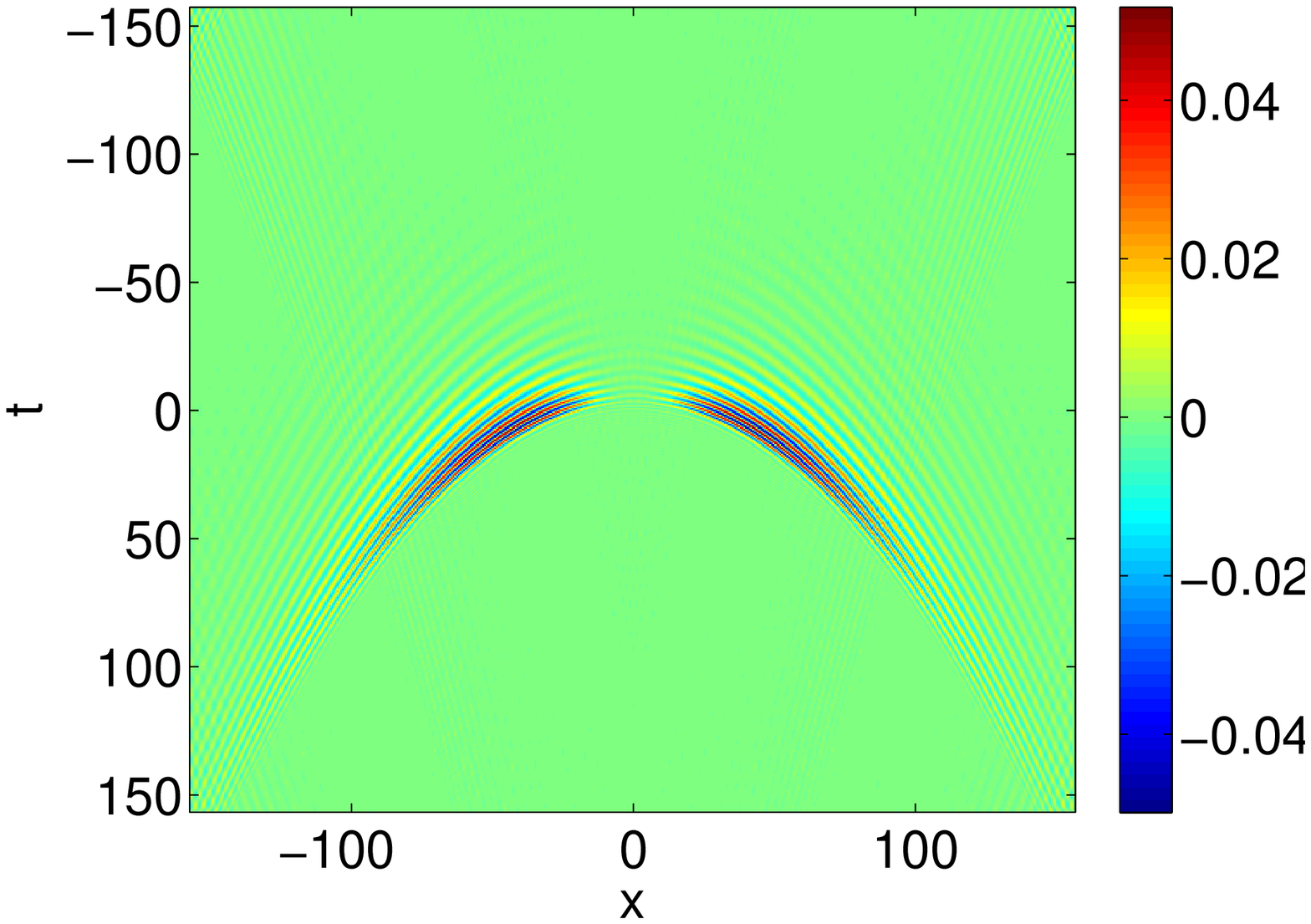}}
{\includegraphics[width=0.22\textwidth]{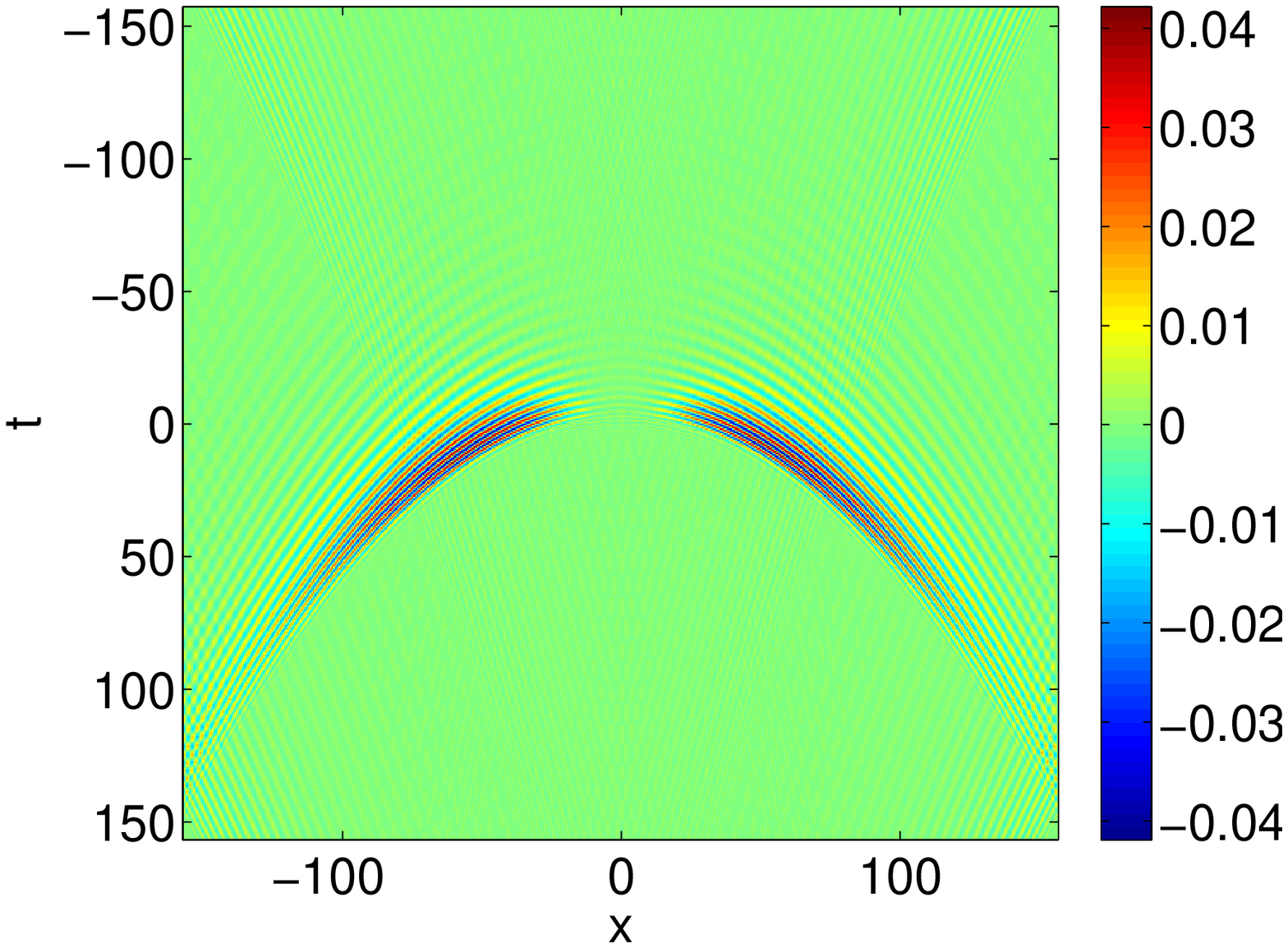}}
\end{center}
\caption{(Color online) Contour plots showing profiles of the field $E$,
in the $(x, t)$ plane, evolved as per SPE-I,
Eq.~(\ref{2dspe_2}), with localized initial data
[cf. Eq.~(\ref{ic1})].
The snapshots, from left to right, and top to bottom
correspond to $z=0,3,10,20,30,38$.
Parameter values are $s=-1/3$ and $\alpha = -2$.
}
\label{localized}
\end{figure}

Taking into  account the above constraint, we now may use
$f(t) = (1-t^2)\exp(-t^2/2)$ which has the above property, and also choose $g(x)$ to be of
the same functional form, namely: $g(x) = (1-x^2)\exp(-x^2/2)$.
The aim of the latter choice is to produce a two-dimensional localized
waveform.
Employing these choices,
we can now numerically integrate Eq.~(\ref{2dspe_2}), using the initial condition:
\begin{equation}
E(z=0,x,t)=(1-x^2)(1-t^2)\exp[-(x^2+t^2)/2].
\label{ic1}
\end{equation}
%
The results of our
simulations are presented in Fig.~\ref{localized}, where we show the evolution
of this initial data. It is clearly observed that, already at
small values of the
propagation distance ($z\approx 2$), the initially localized structure bends and
splits at $(x, t)=(0,0)$, thus forming two ``wing-like'' structures. The size (length) of
these structures is small at the early stages of the evolution but, afterwards, their
spatial extent is increased, as the initial data progressively
disperses. This way, the resulting structures yield, at longer propagation
distances (see bottom panels of Fig.~\ref{localized}),
a quasi-one-dimensional pattern, somewhat reminiscent of the
breather states examined previously.
Here we should mention that our simulations end up at relatively small
distance ($z=38$) in order to avoid interference of these expanding quasi-1D structures with the
boundaries (recall that we use periodic boundary conditions in our numerical scheme).

We also note in passing that we have tried other localized initial conditions, which led
to qualitatively similar results: in all cases, the respective evolutions of
the initial localized
data gradually transformed into quasi-one-dimensional dispersing structures
of the above type.

\section{The SPE-II: general properties and numerical study}

Let us now consider the SPE-II which we express, for simplicity of notation, in the following form:
\begin{equation}
2\sigma(E_{zt}+E_{xt})+\alpha E+s(E^3)_{tt}=0,
\label{2dspe_1}
\end{equation}
where $s= s_{\kappa} \gamma$, as in the SPE-I.
Below we will follow the presentation of the previous section and discuss general properties of this model, such as the corresponding Lagrangian/Hamiltonian
structure and relevant conservation laws
as well as some of its prototypical solutions.

\subsection{Properties and canonical structure}

As in the case of the SPE-I, we integrate Eq.~(\ref{2dspe_1}) with respect to $t$ and, introducing the field
$E=\phi_t$, we express SPE-II in the following form:
\begin{equation}
{2\sigma}(\phi_{xt}+\phi_{zt})+\alpha\phi+s(\phi_t^3)_t=0.
\label{ve_1}
\end{equation}
The above equation can be obtained from the variational principle, with Lagrangian density:
\begin{equation}
\mathcal{L}=-\frac{\alpha}{2}\phi^2+\frac{s}{4}\phi_t^4+\sigma(\phi_t\phi_z+\phi_t\phi_x).
\end{equation}
The corresponding Hamiltonian can then be found as:
\begin{eqnarray}
H&=&\int_{-\infty}^{+\infty} \int_{-\infty}^{+\infty}
\left( \frac{\partial\mathcal{L}}{\partial\phi_z}\phi_z-\mathcal{L}\right){\rm d}t {\rm d}x
\nonumber \\
&=&\int_{-\infty}^{+\infty} \int_{-\infty}^{+\infty}
\left(\frac{\alpha}{2}\phi^2-\frac{s}{4}\phi_t^4-\sigma\phi_t\phi_x \right){\rm d}t {\rm d}x,
\label{H2}
\end{eqnarray}
while the momenta read:
\begin{eqnarray}
M_t&=&\int_{-\infty}^{+\infty}\int_{-\infty}^{+\infty} \frac{\partial\mathcal{L}}{\partial\phi_z}\phi_t{\rm d}t {\rm d}x
=\sigma \int_{-\infty}^{+\infty}\int_{-\infty}^{+\infty}\phi_t^2 {\rm d}t {\rm d}x,
\nonumber \\
\label{Mt2} 
\\
M_x&=&\int_{-\infty}^{+\infty} \int_{-\infty}^{+\infty} \frac{\partial\mathcal{L}}{\partial\phi_z}\phi_x {\rm d}t {\rm d}x
=\sigma \int_{-\infty}^{+\infty}\int_{-\infty}^{+\infty} \phi_t\phi_x {\rm d}t {\rm d}x. \nonumber \\
\label{Mx2}
\end{eqnarray}

Next, we consider the Fourier transform of Eq.~(\ref{2dspe_1}) with respect to $t$, which leads to the equation:
\begin{equation}
2\sigma(i\omega)(\hat{E}_{z}+\hat{E}_x )+\alpha\hat{E}+s\hat{(E^3)}(i\omega)^2=0.
\end{equation}
Solving the above equation with respect to $\hat{E}$ we find:
\begin{equation}
\hat{E}_{z}(\omega,x,z)=-\hat{E}_x-\frac{\alpha}{2\sigma
(i\omega)}\hat{ E}-\frac{s(i\omega)}{2\sigma}\hat{(E^3)}, \ {\rm for} \ \omega \neq\ 0,
\end{equation}
\begin{equation}
\hat{E}(\omega=0,x,z)=0, \ {\rm for} \ \omega \neq\ 0.
\label{omegat=0_12}
\end{equation}
The latter equation leads again to the zero-mass constraint [cf. Eq.~(\ref{constrain})]
that we found in the case of the SPE-I as well. This condition will also be satisfied in our
simulations below.

\subsection{1D breathers and initial data localized in 2D}

We consider traveling wave solutions of Eq.~(\ref{2dspe_1}), in the form $E(\xi,\eta)$,
where the coordinates $\xi$ and $\eta$ are defined as:
\begin{equation}
\xi=z, \qquad \eta=t+cx-cz,
\label{xe}
\end{equation}
where $c$ is an arbitrary real constant. This way, Eq.~(\ref{2dspe_1})
is transformed to the equation:
\begin{equation}
2\sigma E_{\xi\eta}+\alpha E+s(E^3)_{\eta\eta}=0,
\label{speaux}
\end{equation}
which is actually the 1D SPE model \cite{sw}. Since the latter admits breather solutions,
we may follow the procedure described in the previous section and study numerically the
evolution of such a 1D solution in the 2D setting of Eq.~(\ref{2dspe_1}).

\begin{figure}[tbp]
\begin{center}
{\includegraphics[width=0.22\textwidth]{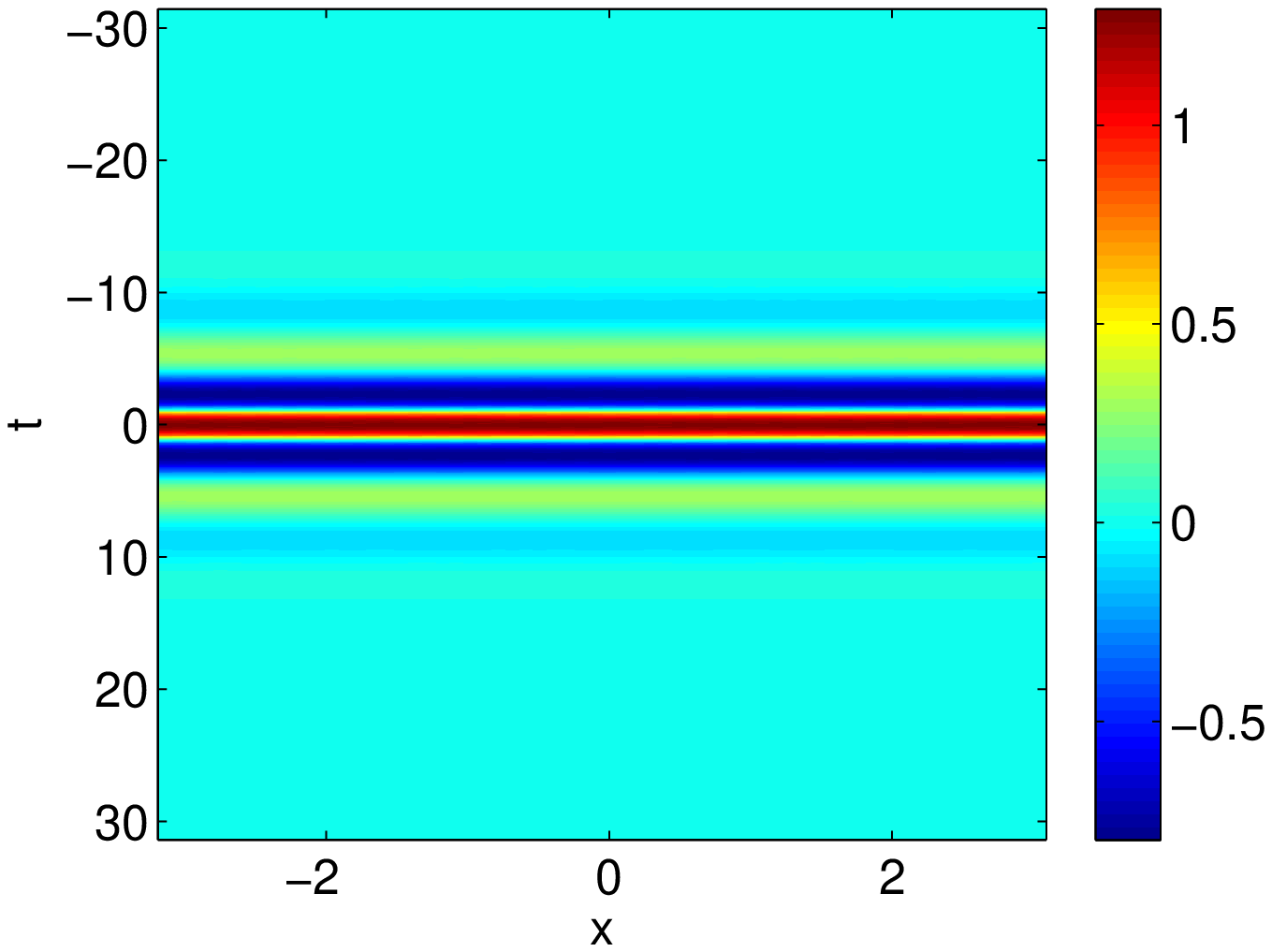}}
{\includegraphics[width=0.22\textwidth]{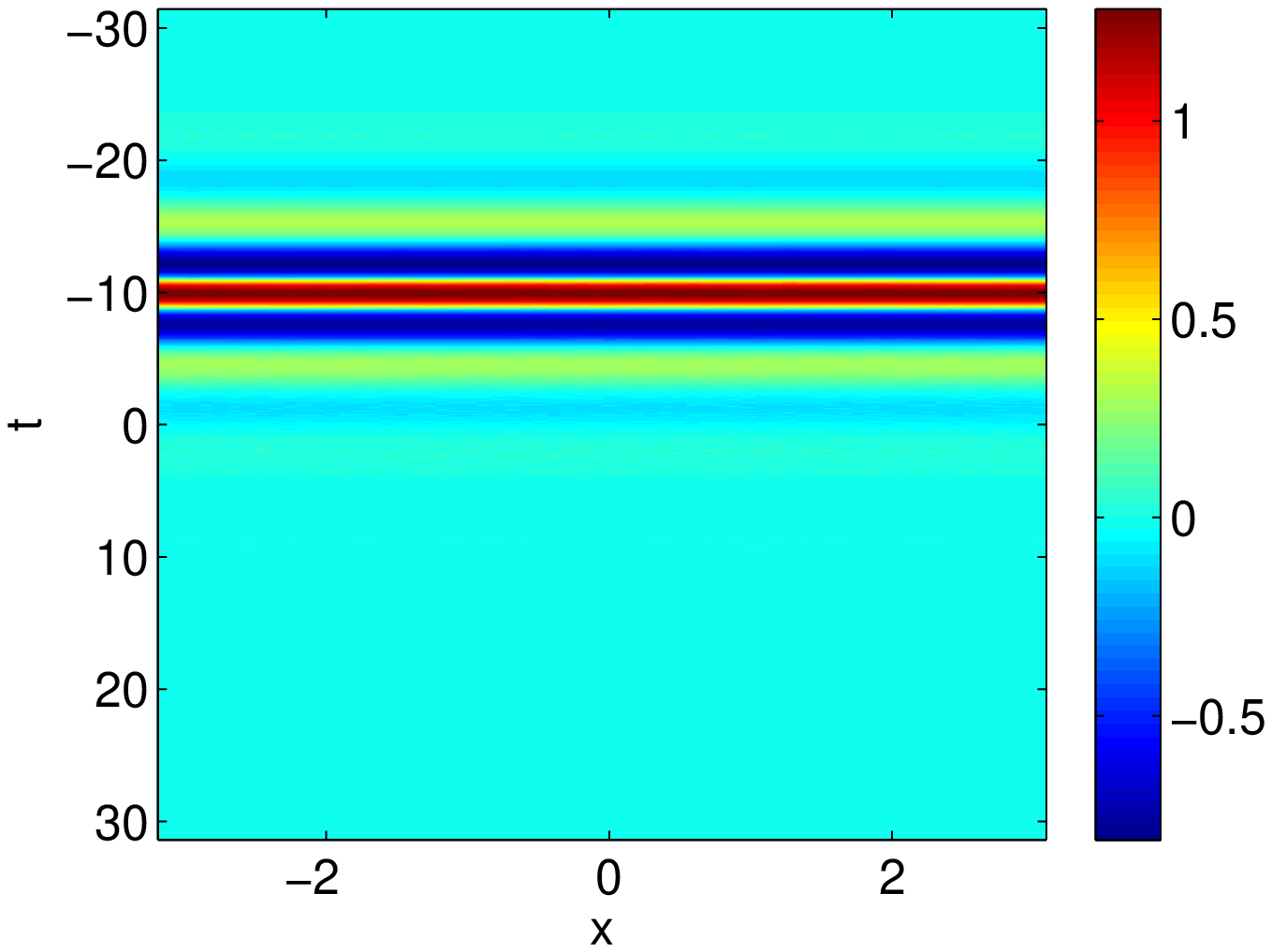}}
{\includegraphics[width=0.22\textwidth]{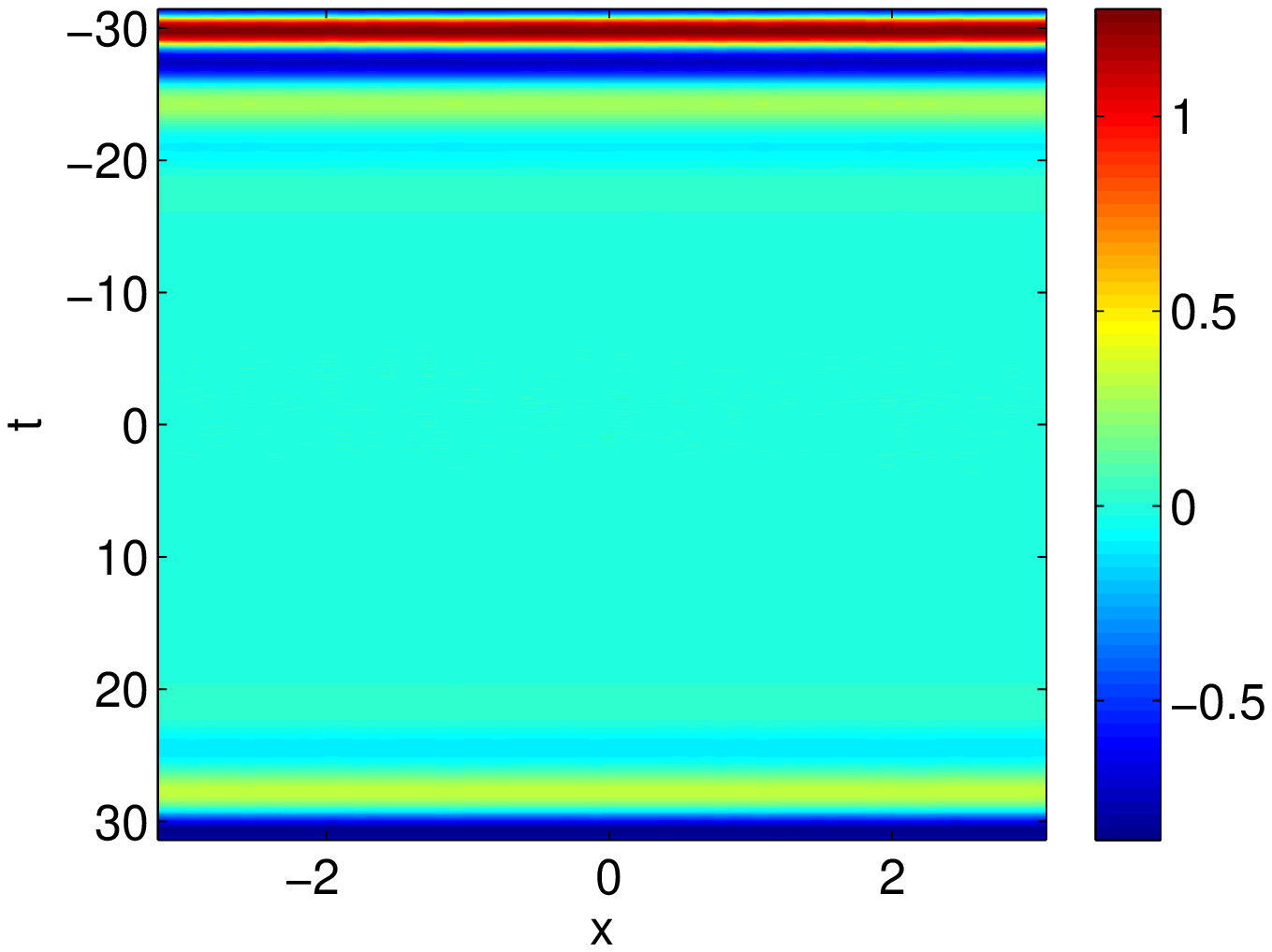}}
{\includegraphics[width=0.22\textwidth]{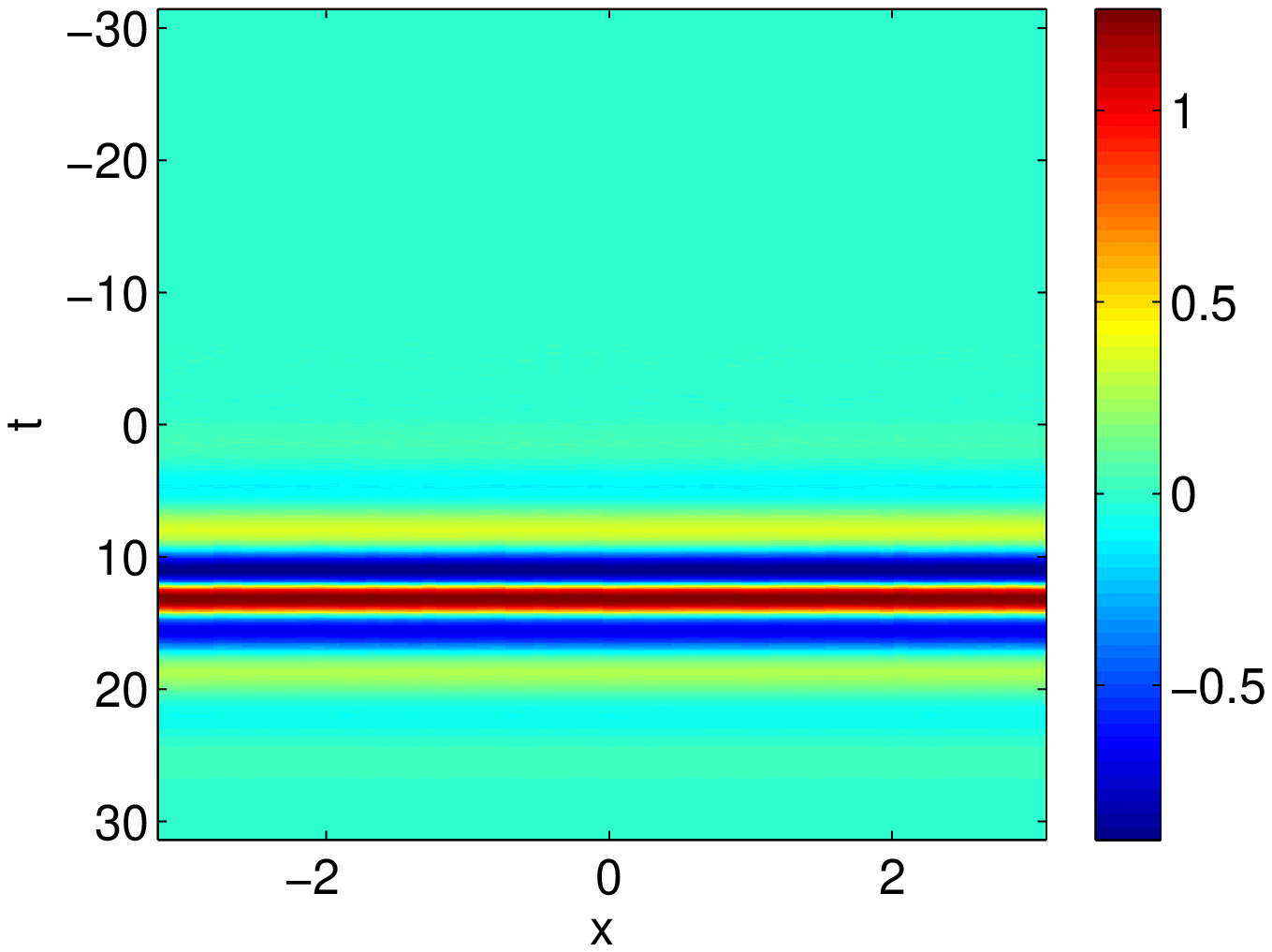}}
{\includegraphics[width=0.35\textwidth]{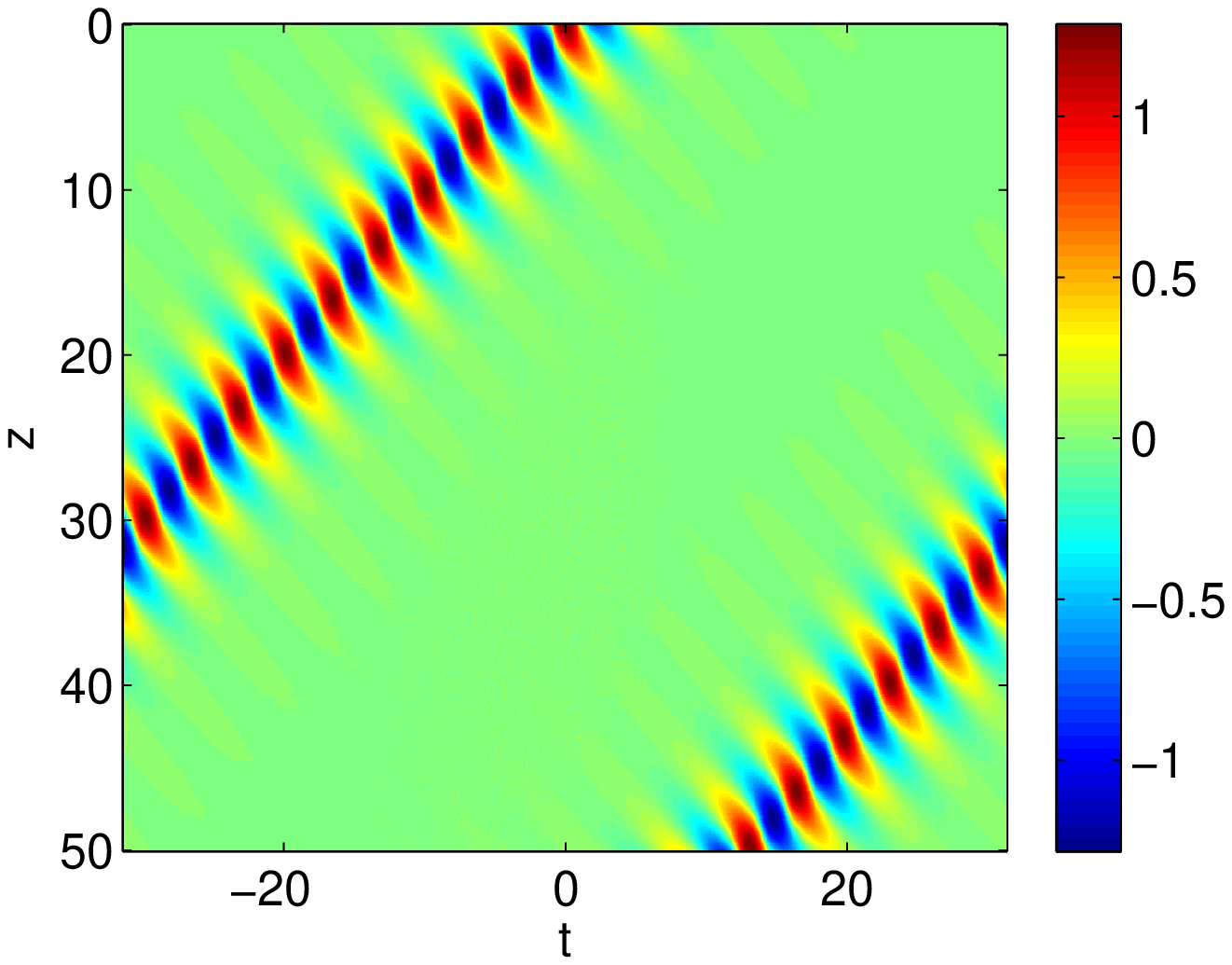}}
\end{center}
\caption{(Color online) Top (four) panels: Contour plots showing profiles of the 1D breather
solution of Eq.~(\ref{speaux}) in the $(x, t$) plane, when evolved as per Eq.~(\ref{2dspe_1}).
Snapshots correspond to $z=0$ (top left), $z=10$ (top right)
$z=30$ (bottom left), and $z=50$ (bottom right).
Bottom panel: The evolution of the breather for $x=0$.
Parameter values are: $c=0$, $s=-1/3$, $\alpha=-2$, and $\sigma = 1$.
}
\label{1Dperturb2}
\end{figure}
\begin{figure}[tbp]
\begin{center}
{\includegraphics[width=0.3\textwidth]{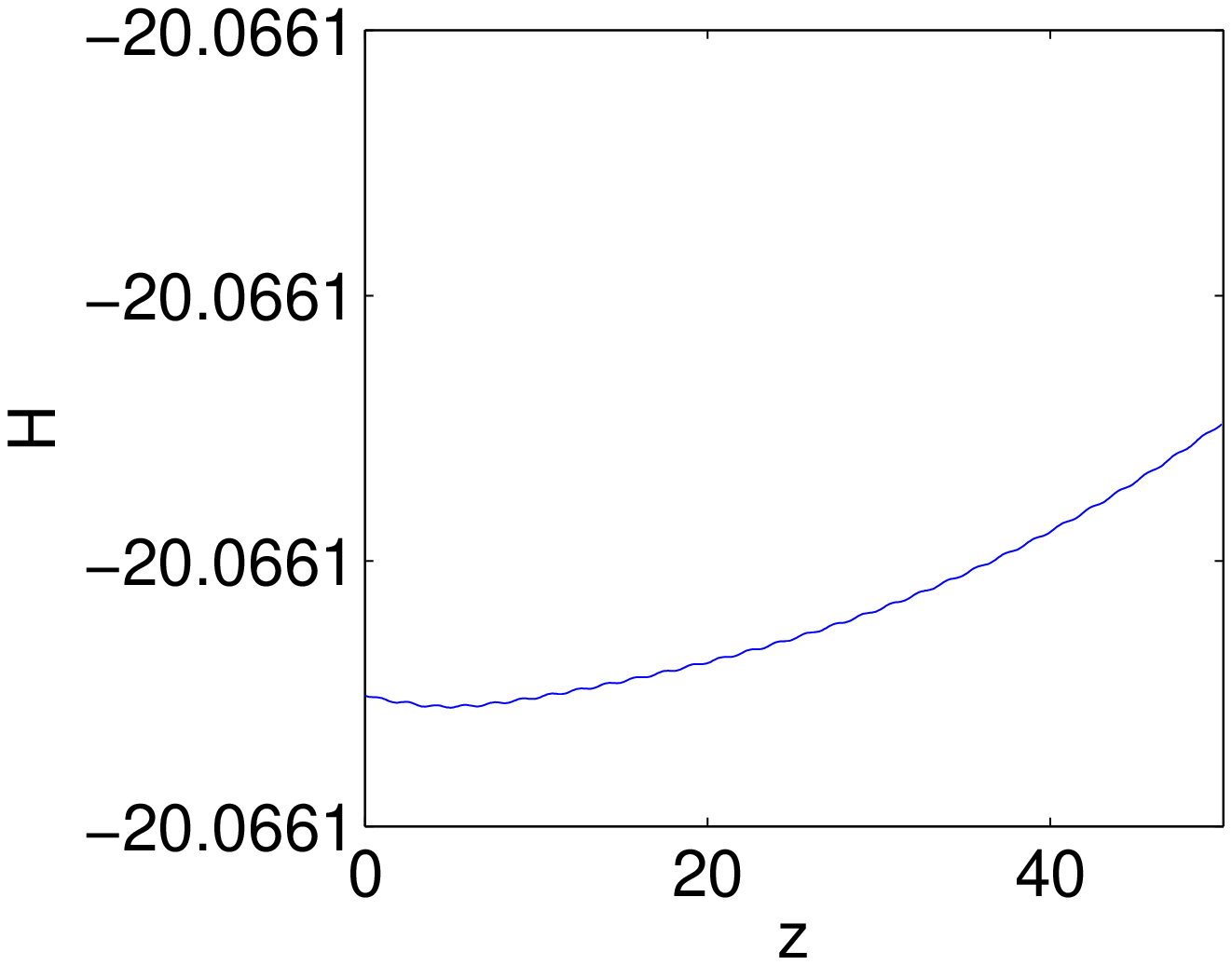}} \\
{\includegraphics[width=0.23\textwidth]{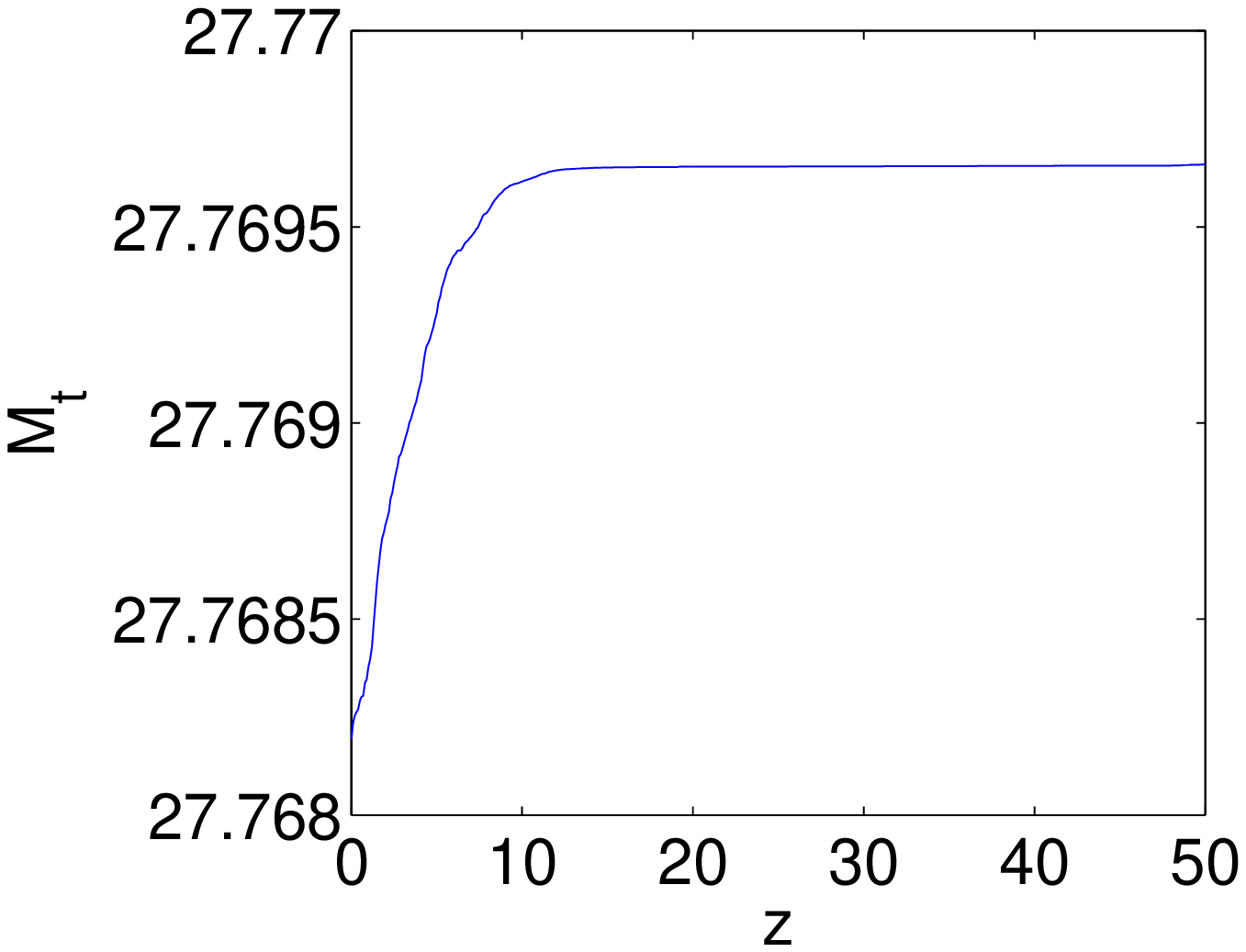}}
{\includegraphics[width=0.23\textwidth]{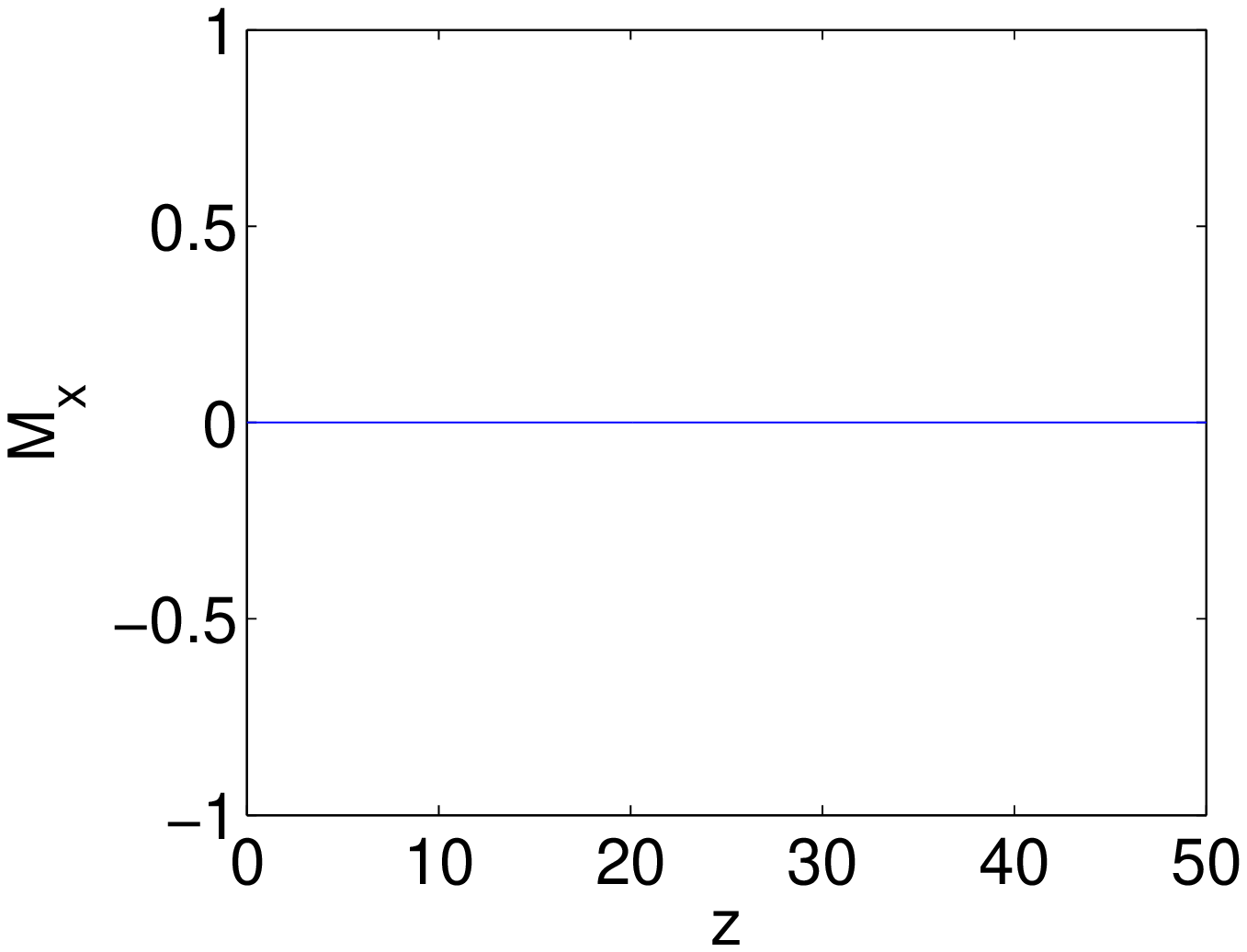}}
\end{center}
\caption{ (Color online) The evolution of the conserved quantities of SPE-II for the simulation shown in Fig.~\ref{1Dperturb2}: the top panel shows the Hamiltonian $H$ and the bottom panels show the momenta $M_t$ (left) and $M_x$ (right). The relative error for $H$ and $M_t$ is of order $10^{-4}$ or less, and the value of $M_x$ is zero (and remains so throughout the simulation).
}
\label{1Dperturb_hamitonian2}
\end{figure}

In this case also, the numerical integration of Eq.~(\ref{2dspe_1}) has shown that this
1D solution is stable in the 2D setting (as was also in the framework of the SPE-I).
An example (pertaining to parameter values $c=0$, $s=-1/3$, $\alpha=-2$, and $\sigma = 1$) is
shown in Fig.~\ref{1Dperturb2}. Additionally, the numerical calculation of the evolution
of the Hamiltonian and momenta [cf. Eqs.~(\ref{H2})-(\ref{Mx2})] depicted in
Fig.~\ref{1Dperturb_hamitonian2}, illustrate
the conservation of these quantities with a relative
error
of order $10^{-4}$ or less. We also note that similar results (not shown here)
were also obtained for oblique moving breathers, i.e., for $c\ne 0$ in Eq.~(\ref{xe}), as in the case of SPE-I.

\begin{figure}[tbp]
\begin{center}
{\includegraphics[width=0.22\textwidth]{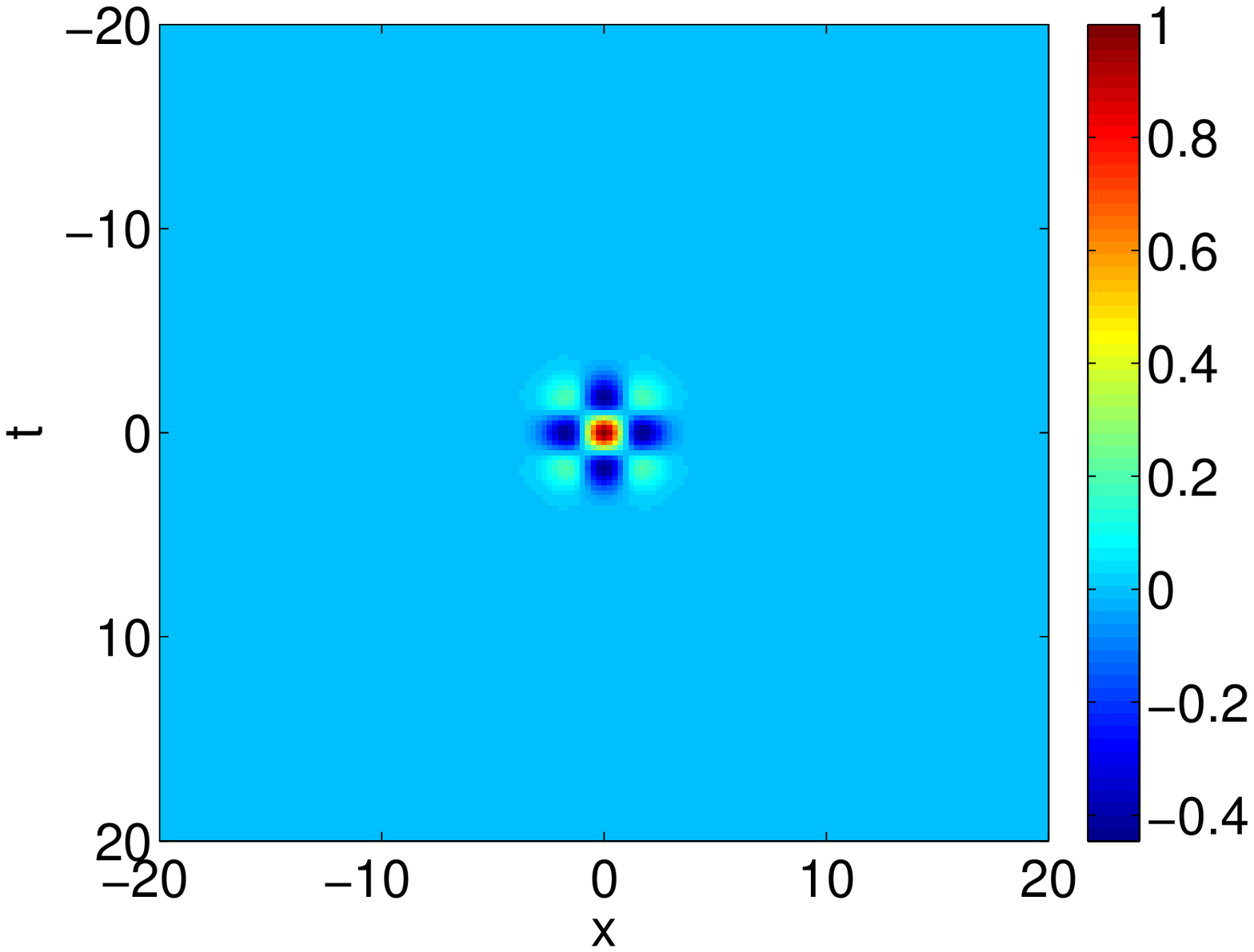}}
{\includegraphics[width=0.22\textwidth]{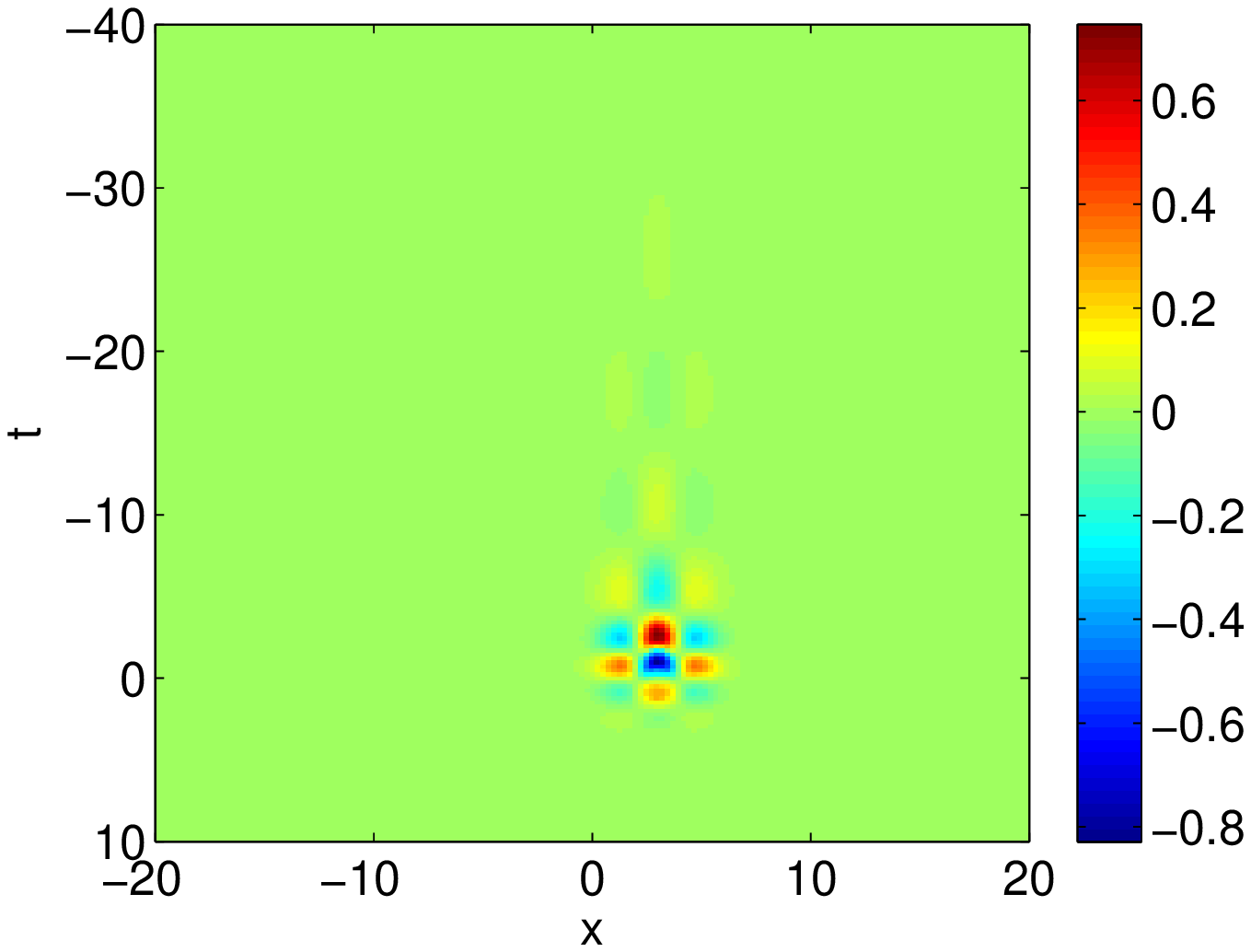}}
{\includegraphics[width=0.22\textwidth]{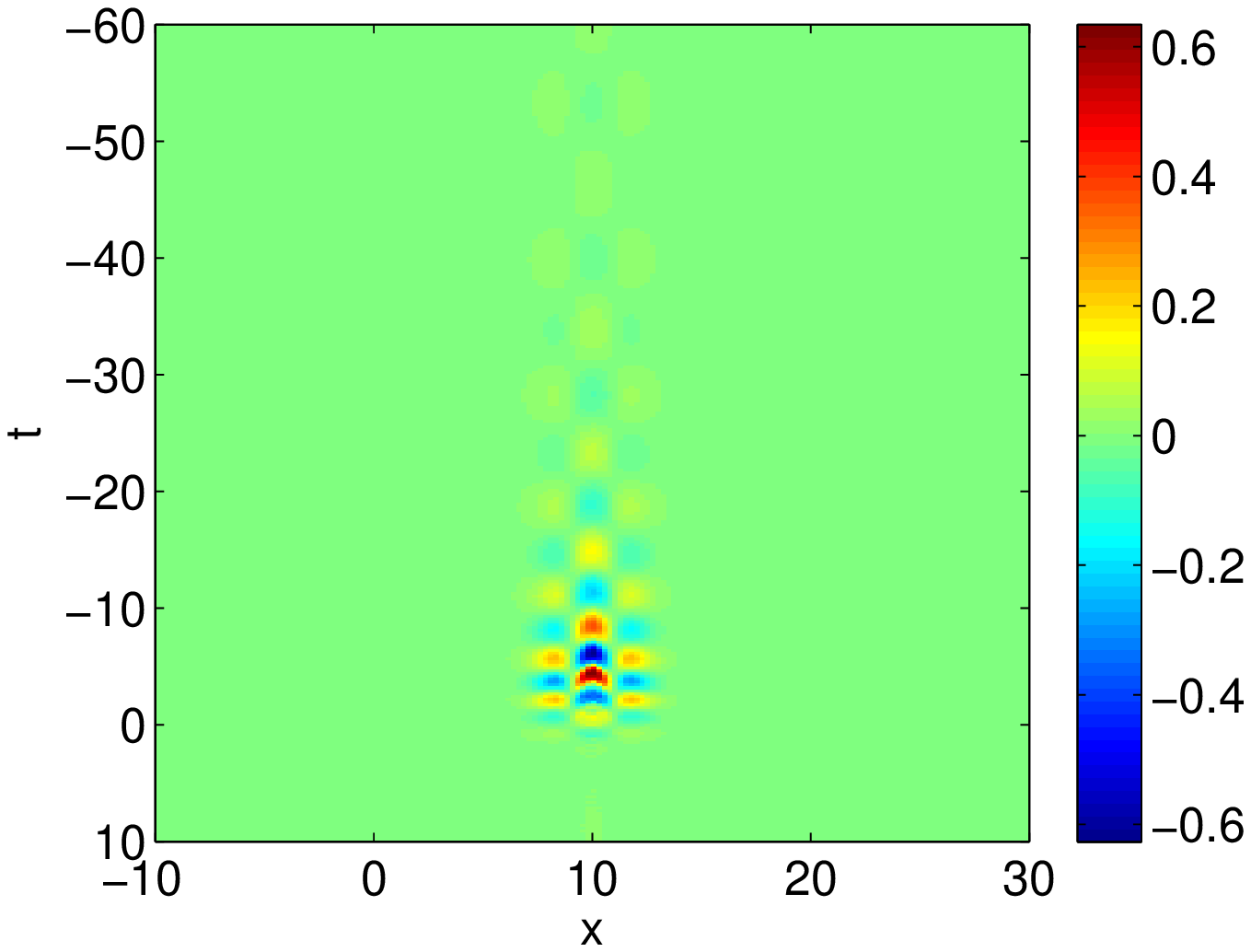}}
{\includegraphics[width=0.22\textwidth]{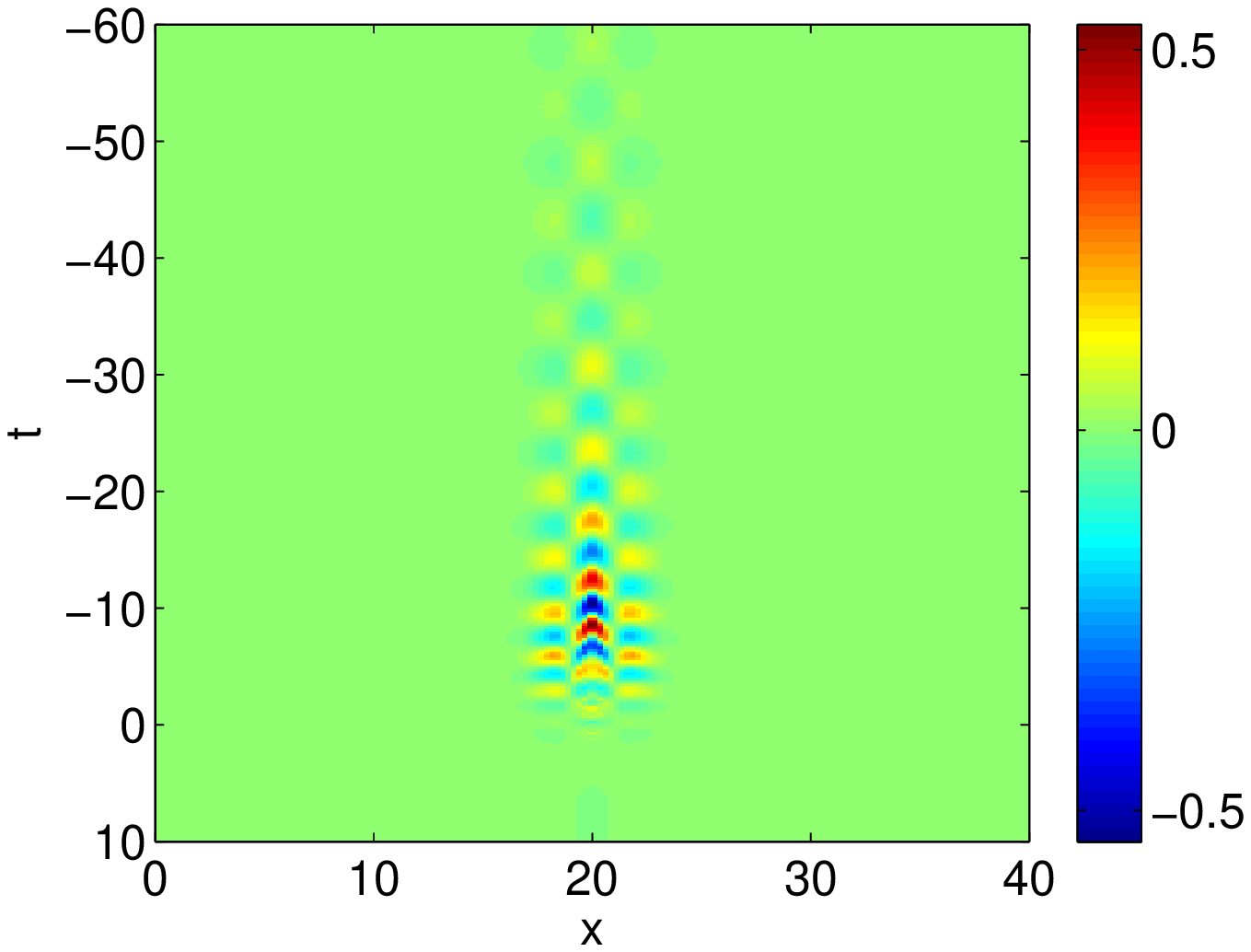}}
{\includegraphics[width=0.22\textwidth]{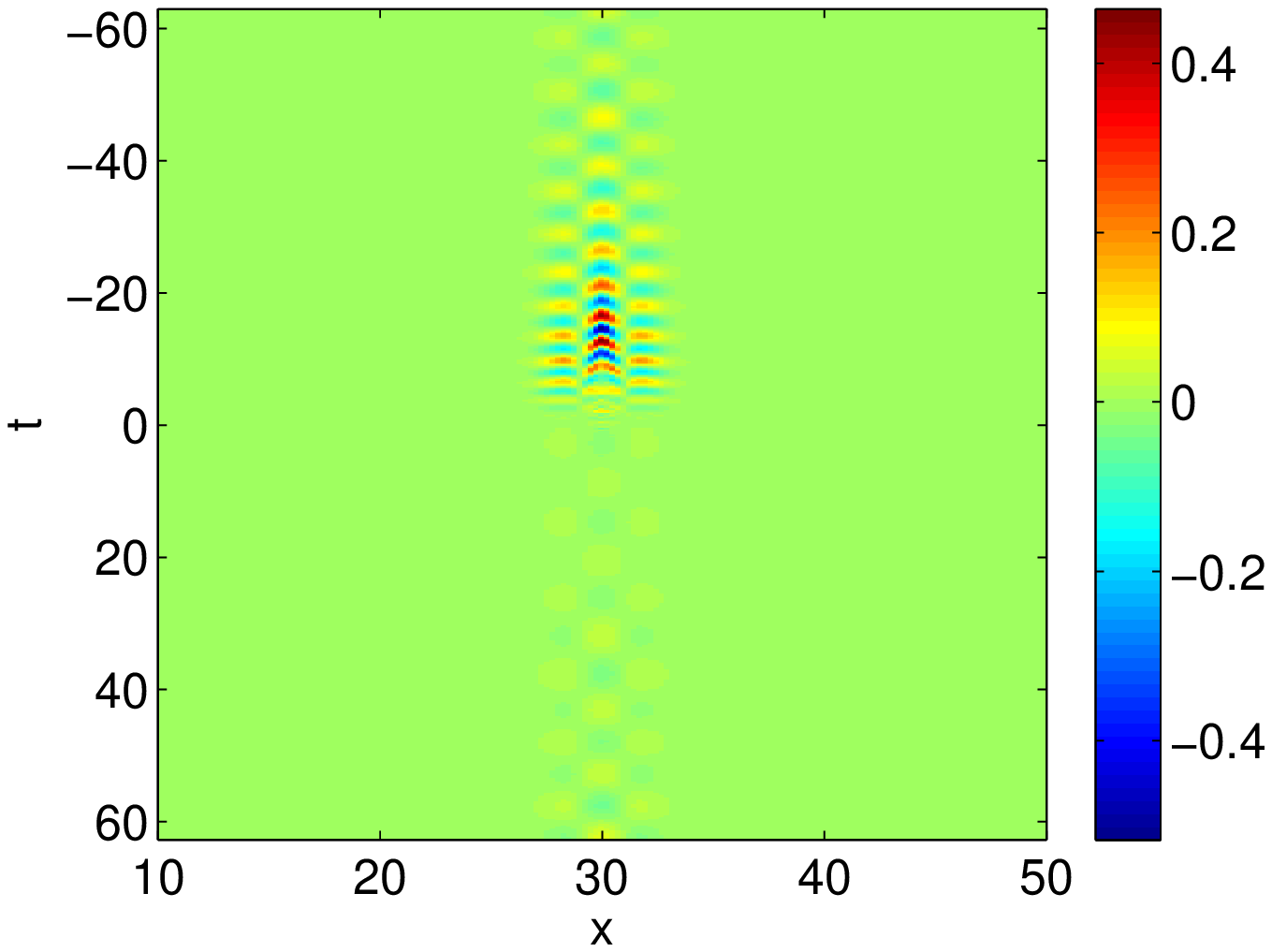}}
{\includegraphics[width=0.22\textwidth]{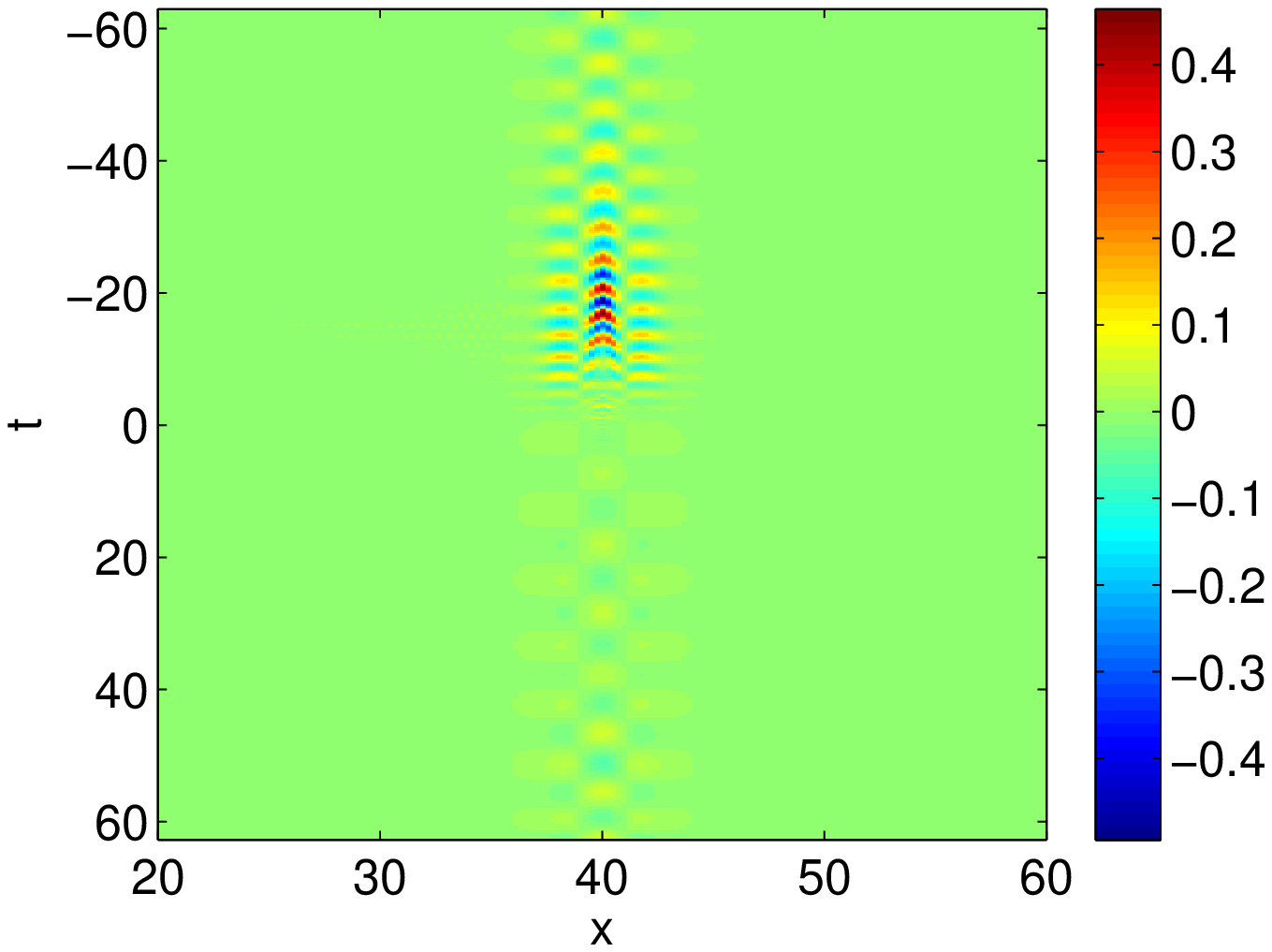}}
\end{center}\caption{
(Color online) Contour plots showing profiles of the field $E$,
in the $(x, t)$ plane, evolved as per SPE-II, Eq.~(\ref{2dspe_2}),
with localized initial data [cf. Eq.~(\ref{ic1})]. The snapshots, from
top to bottom and left to right, correspond to $z=0,3,10,20,30,40$. The domain for numerical computation in $x\times t$ plane is $[-20\pi,20\pi]\times[-20\pi,20\pi]$, we are zooming in here in the snapshots to show more detail.
}
\label{localized_12}
\end{figure}

Finally, as in the case of SPE-I, we study the evolution of localized data in 2D
(i.e., in both $x$ and $t$) in the framework of the SPE-II.
A typical example of the result obtained by the numerical
integration of Eq.~(\ref{2dspe_2}) with such localized initial data
is shown in Fig.~\ref{localized_12}
(parameter values are $s=-1/3$, $\alpha=-2$, and $\sigma = 1$). It is observed that after a small
propagation distance ($z\approx 5$) the initially localized waveform begins to
broaden along the $t$-axis, but still remaining localized along the $x$-axis. As a result,
a quasi-1D structure is gradually formed, which
travels faster along the $x$-direction (where it is localized)
than in the $t$ direction (where it is elongated). In the latter
direction, the structure also possesses an alternating spatial
structure which merits further investigation.

\section{Discussion and Conclusions}

In conclusion, we have derived from Maxwell's equations two $(2+1)$-dimensional short pulse equations,
referred to as SPE-I and SPE-II. These equations may
find applications in various physical contexts where
the study of ultrashort electromagnetic pulses is important; such contexts include nonlinear metamaterials,
nonlinear optical waveguide structures, nonlinear dielectric media, and others. Since both SPE-I and SPE-II actually generalize the $(1+1)$-dimensional SPE \cite{sw}, they can be used to the study of transverse (diffraction-induced) dynamics of ultra-short pulses in such settings. Suitable assumptions
on the nature of the electric and magnetic field and the form of
the permittivity and permeability under which the equations can be derived
were provided.

We have found and presented various general properties of SPE-I and SPE-II. Particularly, we have
identified the Lagrangian and Hamiltonian structure,
and have used invariances to infer (from Noether's theory)
the corresponding momenta, as well as the
associated zero-mass constraints; the latter,
have to be satisfied for the solutions of these equations and, thus,
are also associated with the choice
of the initial data used for
the numerical integration of SPE-I and SPE-II. We have conducted a series
of numerical experiments for
the 2D SPEs using, as initial conditions, either the 1D breather solution
of the underlying
1D SPEs or a localized (in 2D) waveform -- both satisfying the zero-mass
constraint. Our motivation was to
study the stability and transverse dynamics of the  most
robust solutions of the 1D analogue of the system, and also
examine the fate of purely 2D initial data and potentially identify
structures that can be supported by the SPE-I and SPE-II models.

Our numerical simulations have shown that the 1D breathers propagate
(even when they are initially perturbed
by a small noise) practically undistorted. An important conclusion is that
these ultrashort localized
structures are actually insensitive in the presence of diffraction or, in
other words, they appear to be robust in
the presence of (small) transverse perturbations for propagation distances
of the order of a few hundred dimensionless units. On the other hand,
simulations employing initial data localized in
2D have shown that, during evolution, the initial data gradually transforms into quasi-1D structures (which differ between SPE-I and SPE-II).
In fact, we were not able to find any, purely 2D, nonlinear
waveform that can be supported
by either the SPE-I or the SPE-II.

The above results were obtained in the framework of the particular models, i.e., SPE-I and SPE-II, that we
derived and considered in this work. It would be interesting to perform
similar studies (i.e., transverse
dynamics of 1D ultrashort pulses and localized 2D structures) in the
framework of other versions of the
SPE-I and SPE-II, stemming from the incorporation of higher-order effects
(as in the 1D case, in the context of
the so-called regularized SPE \cite{jones,rspe2}). On the other hand, still
in the context of SPE-I and SPE-II, it
would be relevant to consider other types of solutions, e.g., loop-type
solutions or
periodic waveforms composed by
breathers or loops (as in the spirit of the analysis in the 1D case -- see Ref.~\cite{spesol}), and others. It would also be relevant to compare the properties
of the models derived herein with those of other models for ultrashort
pulses including, e.g.,~\cite{koz}-\cite{LKM}. Finally, generalizing the
present models to $(3+1)$-dimensions, removing the assumption of spatial
homogeneity along the $y$-direction would also constitute an interesting
theme for future studies.

{\bf Acknowledgments.} Constructive discussion with  T. P. Horikis are kindly acknowledged.
The work of D.J.F. was partially supported by the Special Account for Research Grants of the
University of Athens. PGK gratefully acknowledges support from the
National Science Foundation under grants DMS-0806762, CMMI-1000337
and from the Alexander von Humboldt Foundation, the Alexander S. Onassis
Public Benefit Foundation and the Binational Science Foundation.


\end{document}